\documentclass{aa} % for the letters 
\usepackage[T1]{fontenc}
\usepackage{epstopdf}
\usepackage{natbib}
\usepackage{graphicx}
\usepackage{txfonts}
   \title{Transient events in bright debris discs: Collisional avalanches revisited}

   \author{P.Thebault
          \inst{1},
          Q.Kral\inst{2}
          }
   \institute{LESIA-Observatoire de Paris, UPMC Univ. Paris 06, Univ. Paris-Diderot, France
    	\and
	   Institute of Astronomy, University of Cambridge, Madingley Road, Cambridge CB3 0HA, UK
             }  
\offprints{P. Thebault} \mail{philippe.thebault@obspm.fr}
\date{Received ; accepted } \titlerunning{Collisional avalanches revisited}
\authorrunning{Thebault \& Kral}

\begin{document}
%

% \abstract{}{}{}{}{} 
% 5 {} token are mandatory
 
  \abstract
   {A collisional avalanche is set off by the breakup of a large planetesimal, releasing vast amounts of small unbound grains that enter a debris disc located further away from the star, triggering there a collisional chain reaction that could potentially create detectable transient structures.}
   {We investigate this mechanism, using for the first time a fully self-consistent code coupling dynamical and collisional evolutions. We also quantify for the first time the photometric evolution of the system and investigate if avalanches could explain the short-term luminosity variations recently observed in some extremely bright debris discs.}
   {We use the state-of-the-art LIDT-DD code (Kral et al., 2013, 2015). We consider an avalanche-favouring A6V star, and two set-ups: a "cold disc" case,  with a dust release at 10\,au and an outer disc extending from 50 to 120\,au, and a "warm disc"  case with the release at 1\,au and a 5-12\,au outer disc.
We explore, in addition, two key parameters, which are the density (parameterized by its optical depth $\tau$) of the main outer disc and the amount of dust released by the initial breakup.}
   {We find that avalanches could leave detectable structures on resolved images, for both "cold" and "warm" disc cases, in discs with $\tau$ of a few $10^{-3}$, provided that large dust masses ($\gtrsim$10$^{20}$--5$\times$10$^{22}$g) are initially released. The integrated photometric excess due to an avalanche is relatively limited, less than 10\% for these released dust masses, peaking in the $\lambda$$\sim$10--20$\mu$m domain and becoming insignificant beyond $\sim$40--50$\mu$m. Contrary to earlier studies, we do not obtain stronger avalanches when increasing $\tau$ to higher values. Likewise, we do not observe a significant luminosity deficit, as compared to the pre-avalanche level, after the passage of the avalanche. These two results concur to make avalanches an unlikely explanation for the sharp luminosity drops observed in some extremely bright debris discs. The ideal configuration for observing an avalanche would be a two-belt structure, with an inner belt (at $\sim$1 or $\sim$10au for the "warm" and "cold" disc cases, respectively) of fractional luminosity $f$$\gtrsim$10$^{-4}$ where breakups of massive planetesimals occur, and a more massive outer belt, with $\tau$ of a few $10^{-3}$, into which the avalanche chain reaction develops and propagates.
   }
   {}

   \maketitle
%
%________________________________________________________________

\section{Introduction} \label{intro}

Debris discs are detected around main sequence stars through the luminosity excess due to circumstellar dust particles. They are believed to be made of the leftover material that has not been used in planetary formation. The standard picture of these systems' evolution is that this material is progressively eroding through a steady-state collisional cascade, starting at the largest bodies in the disc and ending at dust grains small enough to be blown away by radiation pressure \citep{kriv10}.

In addition to being detected in photometry, more than 130 discs have now also been imaged at wavelengths ranging from the optical to the millimeter\footnote{From http://www.astro.uni-jena.de/index.php/theory/catalog-of-resolved-debris-disks.html}. A large fraction of these resolved systems display pronounced spatial structures, such as rings, arcs, spirals, clumps, two-side asymmetries or warps, which bear witness to complex processes shaping these systems \citep[e.g.][]{matthews2014}. Most proposed explanations for these observed structures imply that they are relatively long-lived, being either due to the perturbing effect of (observed or unseen) planets \citep[e.g.][]{moui97,wyat06,reche09, theb12b,nesvold15,2015MNRAS.453.3329P} or stellar companions \citep{theb10,theb12a,nesvold16}, or to interactions with gas \citep{take01,lyra2013}.

\subsection{the collisional avalanche scenario}

\begin{figure}
\includegraphics[scale=0.22]{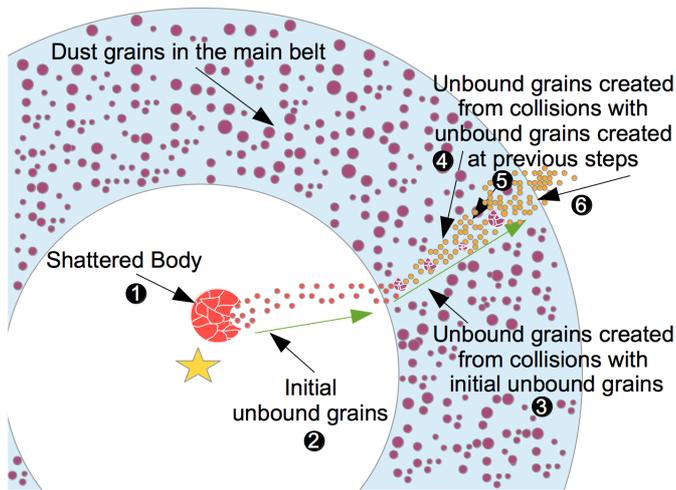}
\caption[]{Schematic description of the avalanche process. First, a large body is shattered in the inner region of the planetary system, releasing very small unbound grains that are blown out by radiation pressure. Then these small grains penetrate an outer disc located further away from the star. There they will collide, at very high velocities, with "native" target grains, producing a new generation of small unbound grains, which will also be blown out and collide with other target grains in the disc, creating new fragments, etc.. If the main disc is dense enough, this chain reaction can produce an excess of small grains largely exceeding the amount initially produced by the breakup.}
\label{sketch}
\end{figure}

Alternatively, some models consider that the observed features are transient and created in the aftermath of the destruction of a large planetesimal or planetary object \citep{jack14,kral15}. One of the main issues with these scenarios is that they require giant impacts that might be uncommon past the first 10--20 Myr of the planet formation phase \citep{wyatt2016}. To alleviate this possible shortcoming, an alternative version of the breakup-aftermath scenario has been proposed, the so-called collisional avalanche mechanism. In this model, first mentioned by \citet{arty97}, the breakup of a planetesimal releases small dust grains below the radiation pressure cutoff size $s_{blow}$, which are placed on outbound trajectories and then impact at very high velocities a debris disc located further out. If this outer disc is dense enough, a chain reaction can be triggered, because these high-$\Delta v$ impacts will produce a new generation of small, $s<s_{blow}$ grains which will again impact particles in the disc, on their way out, producing new small outbound debris, etc (see Fig~\ref{sketch}). This chain reaction might considerably amplify the level of dustiness produced by the initial breakup, thus requiring a less massive object to reach observability than for an isolated planetesimal breakup.

This scenario has first been quantitatively investigated by \citet{grig07} (hereafter GAT07), using a then unique numerical code coupling the dynamical and collisional evolution of a disc. This study has shown that collisional avalanches can indeed strongly amplify the geometrical cross section of the initially breakup-released dust, and produce spiral shaped patterns in the disc that can last for a few orbital periods. GAT07 quantified an avalanche's amplitude by a cross section "amplification factor" $F$ and found that this parameter increases exponentially with the outer disc's density (quantified by its radial optical depth $\tau_{r}$). However, for the nominal case they considered, i.e., a breakup releasing $10^{20}$g of dust and a $\beta$ Pictoris-like disc, GAT07 found that collisional avalanches induce  luminosity increases much too small to be observed. They extrapolated that a system at least 5 times denser than $\beta$ Pic, a disc already at the upper end of bright debris discs, should be required in order for an avalanche to be observable in a \emph{resolved} debris disc. At the time, only one such very bright disc was known, BD+20307 \citet{song05}, an unresolved system, thus not allowing a straightforward application of the avalanche hypothesis.

\subsection{avalanches in "extreme'' debris discs?}

However, in the past decade, several other very luminous discs (with IR fractional luminosities $f \gtrsim 10^{-2}$) have been discovered \citep[e.g.][]{meng2015,genda2015,melis2016}. For these  "extreme" debris discs, the amount of dust required to explain their current luminosities, especially at wavelengths shortward of 24$\mu$m, cannot be sustained by a steady-state collisional cascade over these system's estimated ages \citep[e.g.][]{wyat07b,gasp13,kennedy2013,wyatt2016} \footnote{Note that a debris disc does not necessarily have to be "extremely" luminous in order to exceed the maximum possible luminosity $L_{max}$ expected from a collisional cascade. $L\geq L_{max}$ situations could for example be obtained for lower fractional luminosity $f \sim$10$^{-3}$ values if the system is very old. So these extreme discs do not represent all "anomalously bright" systems as defined by \citet{wyat07b} }. 
The observed dust could be transient, possibly produced in the aftermath of the breakup of a massive body \citep{meng2014,wyatt2016}, or be sustained by comet or asteroid scattering in closely packed multiple planet systems \citep{bonsor2013}.

The discovery of this new class of extreme discs has spawned a renewed interest in the collisional avalanche scenario, which could indeed be greatly facilitated by these system's high level of dustiness. This interest has been compounded by the fact that a handful of these discs exhibit brightness variations on very short timescales, of the order of a year, for which avalanches have been suggested as a possible explanation \citep{meng2015}. It is worth noting that avalanches have been mostly invoked as being able to explain the rapid luminosity \emph{decrease} in some of these systems, in particular the $\sim$50\% decay over $\sim$370 days following the initial brightening of ID8 \citep{meng2014,meng2015}, and the abrupt luminosity drop (by a factor 30) observed for the TYC 8241 2652 1 system over the course of less than two years \citep{melis2012}. The argument is that the passage of the avalanche would leave the main disc depleted in $\lesssim 100\mu$m particles, as they have been "sandblasted" by the high-velocity impacts with the outbound avalanche grains, making the post-avalanche system (after the transient luminosity excess due to unbound grains) less luminous than it was prior to the avalanche.
Interestingly, this luminosity drop consequence of an avalanche had not been mentioned by GAT07, who insisted only on the luminosity \emph{increase} that it might trigger. It is true that the temporal evolution of the amplification factor $F$ showed a progressive decay after a strong initial increase, but this parameter only measures the avalanche geometrical cross section with respect to that of the initially released grains, but \emph{not} that of the total avalanche+disc system. As a matter of fact, the photometric evolution of the total system was not investigated in GAT07, the focus being essentially on spatial signatures and localized luminosity contrasts.

\subsection{revisiting the avalanche mechanism}

In the light of this new context, it appears important to reinvestigate the collisional avalanche mechanism and reassess its relevance with respect to observed systems. Such a reinvestigation was in any case necessary and long overdue, firstly because the pioneering study of GAT07 left many issues unexplored (photometric evolution, wavelength dependence, etc.), which maybe opened the door to some misconceptions about the avalanche phenomenon, but also because new debris disc modelling codes have been developed in the past decade, which allow a more accurate study of this complex mechanism. The most advanced of this new generation of codes is probably the LIDT-DD model developed by \citet{kral13}, which is to date the only that allows for a coupled study of collisions and dynamics taking into account the crucial size-dependent effect of radiation pressure. LIDT-DD allows to relax some of the simplifications done in GAT07, mainly the fact that "avalanche" (i.e., the initially released debris and all new debris they will collisionally generate) and "field" (i.e., grains in the outer disc not affected by collisions with avalanche grains) particles were treated separately, but also that the possible presence of "native" unbound grains in the field disc itself was not taken into account.

We thus here revisit the collisional avalanche scenario using the state-of-the-art LIDT-DD code. Furthermore, we now consider two different configurations: a "cold "disc" case similar to that of GAT07, with the breakup event at 10\,au and the outer disc at 50-120\,au, but also a "warm disc" case, closer to the new class of observed "extreme" discs, with the breakup at 1\,au and the outer disc at 5-12\,au, and explore different values of the main disc density.
We present the main characteristics of our model and set-up in Section 2. Results for all explored cases are presented in Section 3. In Section 4, we discuss the main implications of these results, in particular the dependence of avalanche strength on disc density and the likelihood of avalanche-triggering events in real systems. A detailed summary and our conclusions are given in the last section.

\section{Model}\label{model}

\subsection{the LIDT-DD code}

The LIDT-DD code has been described in detail in \citet{kral13,kral15}, but we recall here some of its main characteristics. It is in essence a hybrid code, taking some of its structure from a protoplanetary-disc model developed by \citet{char12}, coupling an $N$-body scheme to compute the dynamics and a statistical particle-in-a-box approach to follow the collisional evolution. The base elements of this model are "super particles" (SPs), each representing a cloud of grains of a given size (or, more exactly, a size bin) at a given location, whose dynamical evolution is computed with a standard Burlish-Stoer algorithm. The collisional evolution is then performed by dividing the system into 3D spatial cells, and estimating all mutual collisions between the SPs contained in each cell. Each of these mutual SP collisions is treated in a statistical way, as being between two clouds of grains having the physical sizes represented by each SP and sharing their local dynamical characteristics. The physical outcomes of the grain-grain collisions can be fragmentation, cratering or accretion, depending on the grain sizes and relative velocities.
The collisional fragments produced by cratering or fragmenting collisions will either be reattributed to existing SPs in the same cell, or assigned to newly created SPs in case SPs representing similar sizes and dynamical characteristics do not already exist in the cell.

One crucial aspect of LIDT-DD, which sets it apart from other similar hybrid codes \citep[e.g.,][]{levison12,nesvold13}, is that it takes into account the size-dependent effect of stellar radiation pressure. Because dynamical evolution then strongly varies with the grains' physical sizes, each SP only stands for one given size bin, instead of representing a whole size distribution as in other codes. Given the wide radial excursion of small grains affected by radiation pressure, this does also mean that, in any given region of the system, SPs representing the same grain size can have very different dynamical characteristics depending on where the grain population they stand for has been originally produced. Accordingly, the code is equipped with a procedure that sorts out, for any given size bin in a given spatial cell, all SPs into distinct dynamical families that collisionally interact with each other.

\subsection{Setup}\label{setup}

In accordance with the standard avalanche scenario, we consider here the breakup of a planetesimal, producing a cloud of fragments, of which a fraction are placed on unbound orbits and will impact grains located in an extended main debris disc located further out (Fig.~\ref{sketch}).   

We consider two basic configurations. One of them is the same as in GAT07, with an initial breakup occurring at 10\,au from the central star, and the outer disc located between 50 and 120\,au. This "cold" debris disc case is considered essentially in order to investigate the spatial signature of an avalanche, because this $\sim$50--100\,au range is the one where most spatial structures have been observed in imaged debris discs \citep{2014AJ....148...59S}.
The other configuration is scaled down by a factor of 10, with a breakup occurring at 1\,au and  the outer disc between 5 and 12\,au. This "warm disc" case is motivated by the study of the potential photometric signature of an avalanche, in particular in the $\lesssim24\mu$m wavelength domain where most of the short term luminosity variations of "extreme" discs have been observed.

\begin{figure}
\includegraphics[scale=0.35]{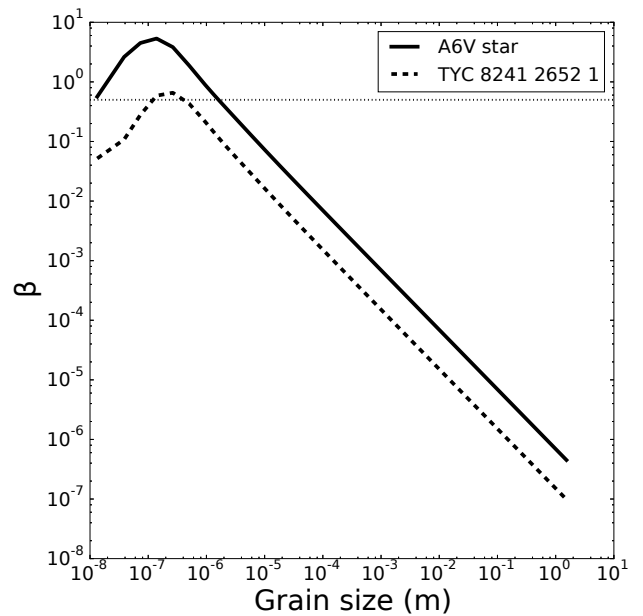}
\caption[]{Value of $\beta$, the ratio between radiation pressure and gravitational forces, as a function of particle size, computed for non-porous astrosilicates. The $\beta$(s) curve is shown for the luminous A6V star considered as a reference case in our runs, as well as for the TYC 8241 2652 1 K star discussed in Sec.\ref{extreme}. The dotted horizontal line marks the $\beta=0.5$ limit, above which grains are no longer on bound orbits around the star (if produced from progenitors on circular orbits)}
\label{betafig}
\end{figure}

In order to avoid multiplying the number of free parameters to an unmanageable level, some of them are fixed in our set-up. For the grain composition, we consider a standard astrosilicate case \citep{drai03} with zero porosity. Classically, the grain's resistance to impacts is parameterized by the critical specific energy $Q^*$ required to disperse 50\% of an impacted target, for which we assume the empirical expression derived by \citet{benz99}. Collision outcomes (cratering, fragmentation) are then treated following the procedure described in \citet{kral13}. 
For the central star, we assume an A6V spectral type, similar to the $\beta$-Pic case considered in GAT07, as such a luminous early-type star is more favourable to avalanche development. Indeed, for such stars, there is a much larger fraction of dust grains with $\beta >0.5$ ($\beta$ being the ratio between stellar radiation and gravity, and $\beta$=0.5 being the limit for unbound orbits for grains produced from circular orbits) than for solar-type stars \citep[see for instance Fig.~2 in][]{2001A&A...366..945B}. The initial eccentricities of particles in the main disc are uniformly distributed between 0 and $e_{max}=0.1$ and their inclinations between 0 and $e_{max}/2$, which is a standard assumption for "classical" debris discs \citep[e.g.][]{theb09}.
We follow the evolution of all particles with sizes comprised between $s_{min}=$\,20nm ($2\times10^{-8}$m) and $s_{max}=$ 1m, distributed onto 48 logarithmically spaced size bins. Note that $s_{min}$ is well below the blowout size for an A6V star, which is $s_{blow}\sim$1.6$\mu$m for compact astrosilicates, so that we cover a large size range of bodies with unbound orbits ($\beta\geq0.5$, see Fig.\ref{betafig}).
The initial size distribution follows a standard power law in $dN\propto s^{-3.5}ds$ down to the cutoff size $s_{blow}$, but this initial profile is quickly relaxed as we let the system evolve towards a collisional steady-state before launching the avalanche.

This leaves us with two main free parameters, both crucial to avalanche development. The first one is of course the mass $m_{\rm{init}}$ of dust released by the initial planetesimal breakup. GAT07 considered a case with $m_{\rm{init}}=10^{20}$g of grains of size $s<1$cm, corresponding to the mass of an object of radius $\sim$20\,km, and we explore different values of $m_{\rm{init}}$ around this reference value. Of course, not all of the fragments produced by this hypothetical shattered planetesimal of mass $M_{breakup}$ are in the $s<1\,$cm range, but we defer a discussion about the link between $m_{\rm{init}}$ and $M_{breakup}$ to the discussion. The size distribution of the released fragments follows a slightly steeper law in $dN\propto s^{-3.8}ds$, corresponding to the crushing law expected for the outcome of violent massive collisions \citep[e.g.][]{taka11,lein12}, all the way down to $s_{min}=2\times10^{-8}$m (see Sec.\ref{limit} for a discussion on this point).

The second free parameter, which GAT07 considered to be the most important one, is the dust density in the main outer disc. GAT07 parameterized this density by the radial optical depth $\tau_{r}$, which scales the amount of material an outward moving grain will cross on its way through the disc. However, we choose here to take the more commonly used normal optical depth parameter $\tau_{\perp}$. This parameter then allows comparison to one easily observable quantity  that is the fractional infrared luminosity $f=L_{IR}/L_{bol}$ \citep{wyat07a}, at least as long as the grains dominating the geometrical cross section excess do also dominate the excess in IR flux (which might not always be the case, see discussion in Sec.\ref{gat07}). For a disc of radial width $\Delta r$ located at an average distance $r$ from its star with an average normal optical depth $\tau_{\perp}$, we have \citep{zuck12}:
\begin{equation}
\tau_{\perp} = 2\,f\,\frac{r}{\Delta r}
\end{equation}
And the relation between $\tau_{\perp}$ and $\tau_{r}$ is
\begin{equation}\label{tau}
\tau_{r} = \tau_{\perp}\left(\frac{\Delta r}{h}\right),
\end{equation}
where $h$ is the vertical thickness of the disc. For the considered case of an extended disc with $\Delta r/r\sim$0.8, $\tau_{\perp}$ and $f$ only differ by a factor 1.5 \footnote{The fractional infrared luminosity $f$ would be equal to the radial optical depth if the dust was distributed onto a spherically symmetric shell \citep{zuck12}}, whereas the ratio between $\tau_{r}$ and $\tau_{\perp}$ is $\sim$5--10 if we consider a typical disc thickness of 0.05--0.1 \citep{theb09}. We consider as a reference the $\left<\tau_{\perp}\right> \sim$2$\times10^{-3}$ value (and thus $\tau_{r}\sim$0.02) taken by GAT07 for their $\beta$-Pictoris-like case, and explore values around this nominal case.
As for the radial distribution of the particle volumic number density within the main disc, we assume that it follows an initial profile in $1/r$, corresponding to a flat radial distribution for $\tau_{\perp}$.
The setup is summarized in Table~\ref{truc}.

The procedure for the simulations is then the following. We first consider the main outer debris disc, which we let evolve until a collisional steady state is reached, i.e., until the \emph{shape} of the size distribution no longer evolves and only its absolute level decreases. We then release the avalanche dust grains and follow the evolution of the combined avalanche+main-disc system. Contrary to GAT07, we do not measure the amplitude of an avalanche by the cross section amplification factor $F$, because this parameter, while allowing a good description of how an avalanche unfolds, does not give any indication about the real signature it leaves in a disc and its observability. We instead adopt a more classical approach where we evaluate an avalanche's strength by two criteria: firstly by estimating, on 2D synthetic images, the luminosity contrast of avalanche-related structures with respect to the background disc, and secondly by measuring the system-integrated photometric variations it induces. For deriving both these synthetic images and the integrated fluxes, we use the GRaTer radiative transfer package \citep{auge99}.

\begin{table}[h]
\caption{Set-up for the "cold" and "warm" disc configurations}
\centering
\begin{tabular}{lc}
\hline\hline
Star\\
\hline
Stellar-type & A6V \\
Mass        & $1.84 \, M_\odot$\\
\hline
Grain physical characteristics\\
\hline
Material & Silicate \\
Blow-out size ($s_\mathrm{blow}$) & $1.6 \, \mu m$ \\
Porosity & 0 \\
\hline
Main outer disc \\
\,(\textit{then evolves towards collisional steady state})\\
\hline
Minimum simulated dust size & $ 0.02\,\mu$m  \\
Maximum simulated dust size   & 1 m  \\
Initial size distribution & $\textrm{d}N \propto s^{-3.5} \textrm{d}s$ down to $s_\mathrm{blow}$\\
Initial radial extension & 50-120\,au (cold disc case) \\
  & 5-12\,au (warm disc case) \\
Average vertical optical depth & $\tau_{\perp}$$\sim$2$\times 10^{-3}$ (nominal case)\\
 & $\tau_{\perp}$$\sim$10$^{-2}$ (dense disc)\\
 \hline
Post-breakup released dust\\
\hline
Minimum size     & $ 0.02\,\mu$m  \\
Maximum size   & 1 cm  \\
Initial size distribution & $\textrm{d}N \propto s^{-3.8} \textrm{d}s$ \\
Initial total mass $m_{\rm{init}}$& $5\times10^{22}$ g (cold disc case)\\
  & $10^{20}$ g (warm disc case)\\
Release distance from the star & 10\,au (cold disc case) \\
& 1\,au (warm disc case) \\
\hline
\end{tabular}
\label{truc}
\end{table}

\section{results}\label{hdresults}

\begin{figure*}
\makebox[\textwidth]{
\includegraphics[scale=0.5]{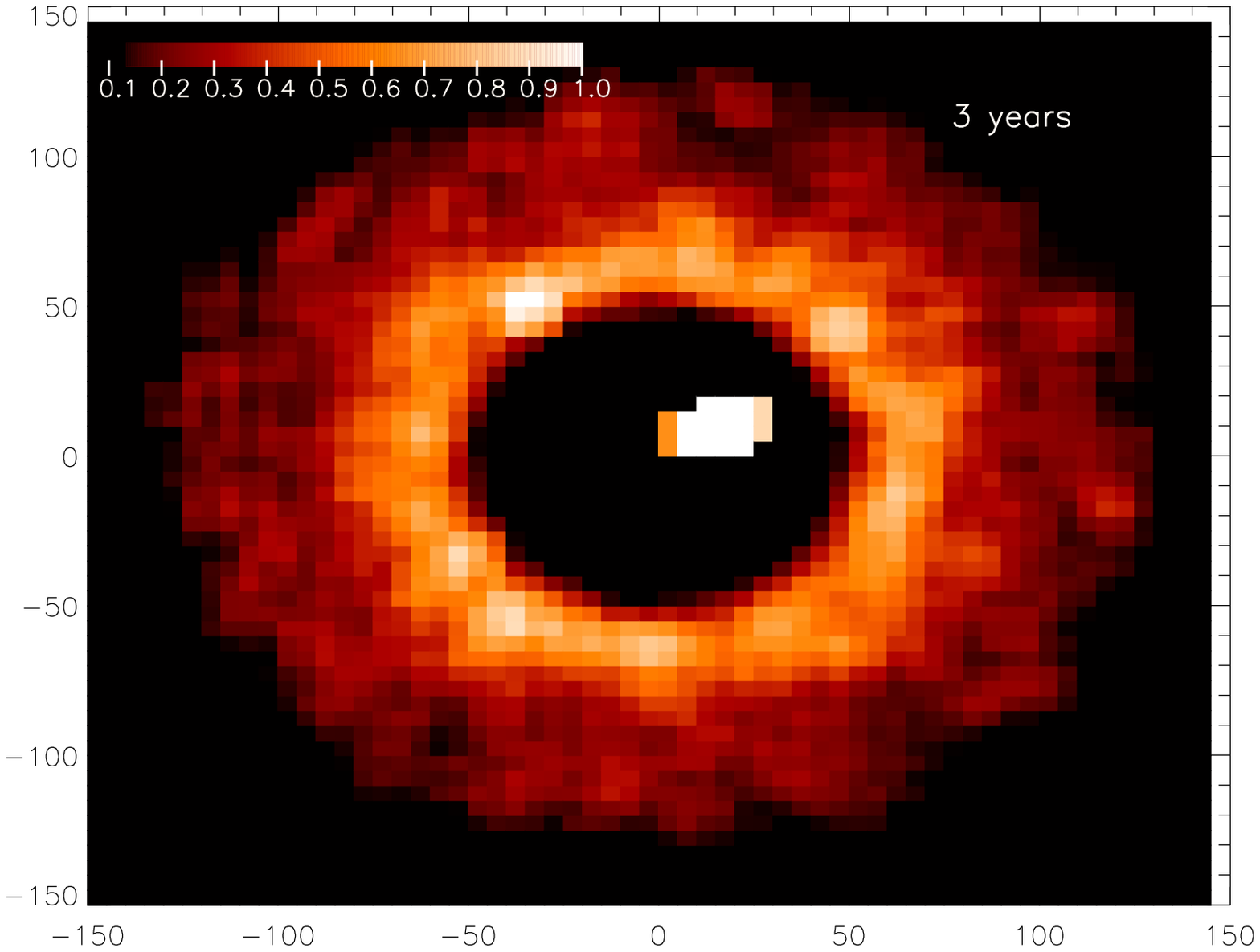}
\includegraphics[scale=0.5]{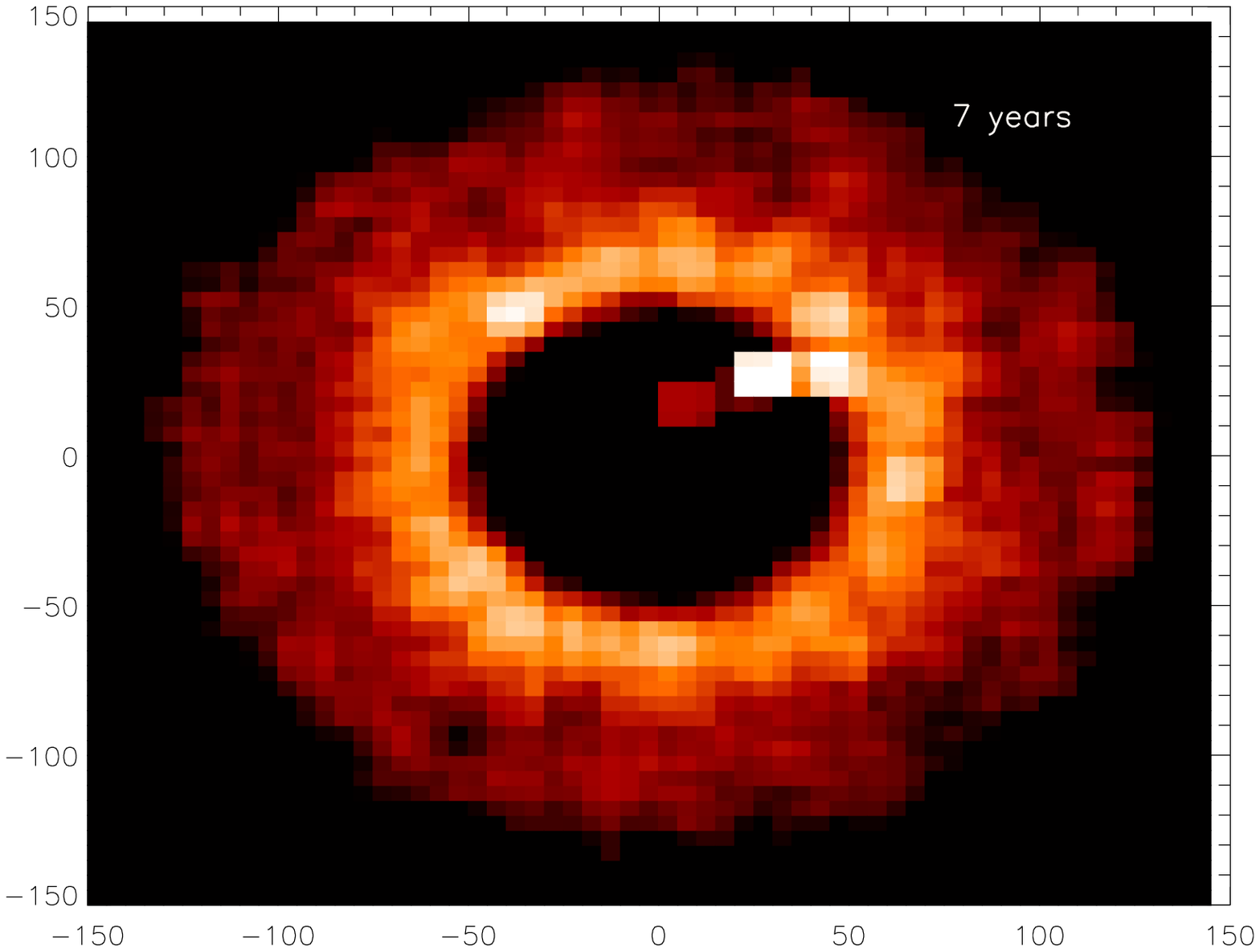}
}
\makebox[\textwidth]{
\includegraphics[scale=0.5]{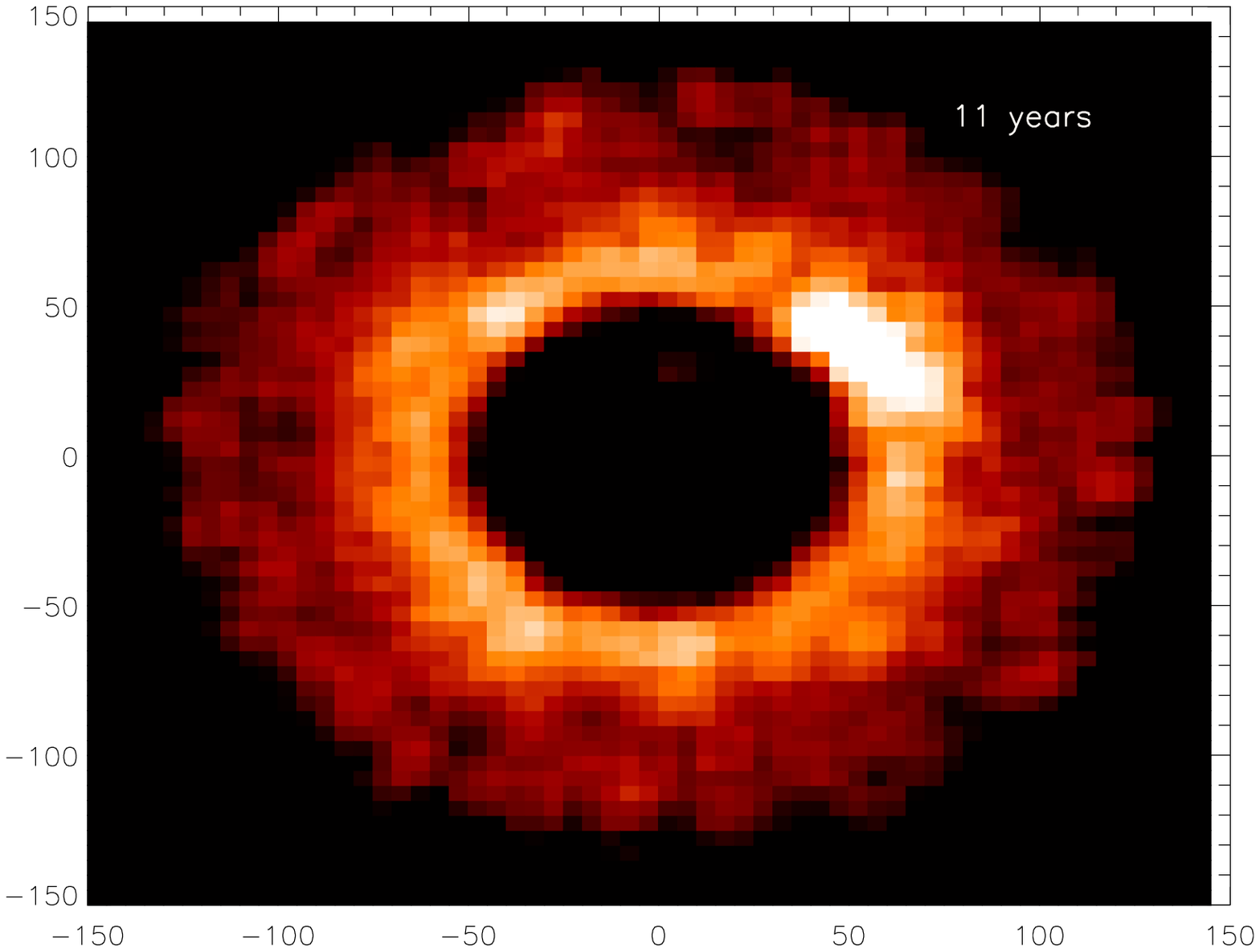}
\includegraphics[scale=0.5]{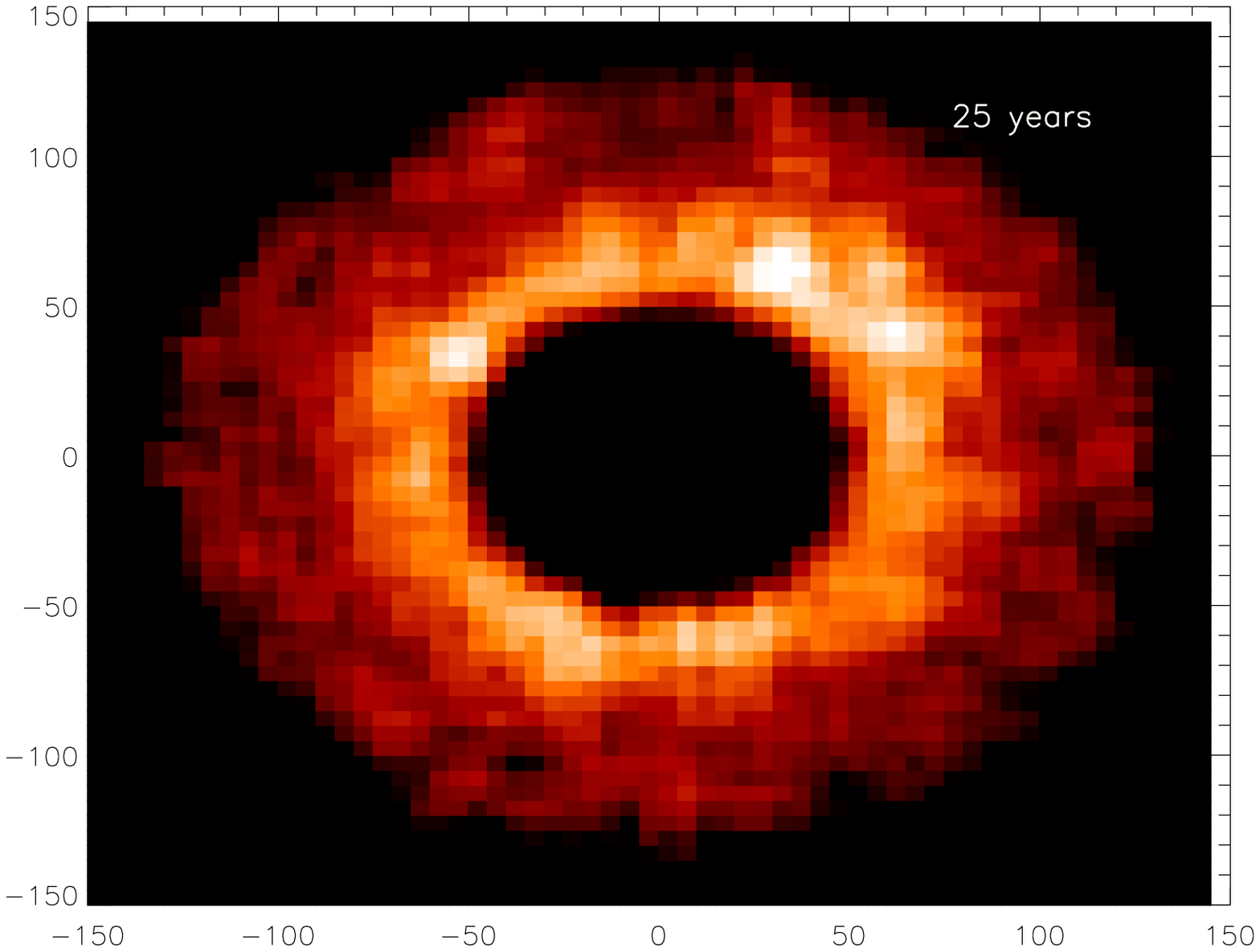}
}
\makebox[\textwidth]{
\includegraphics[scale=0.5]{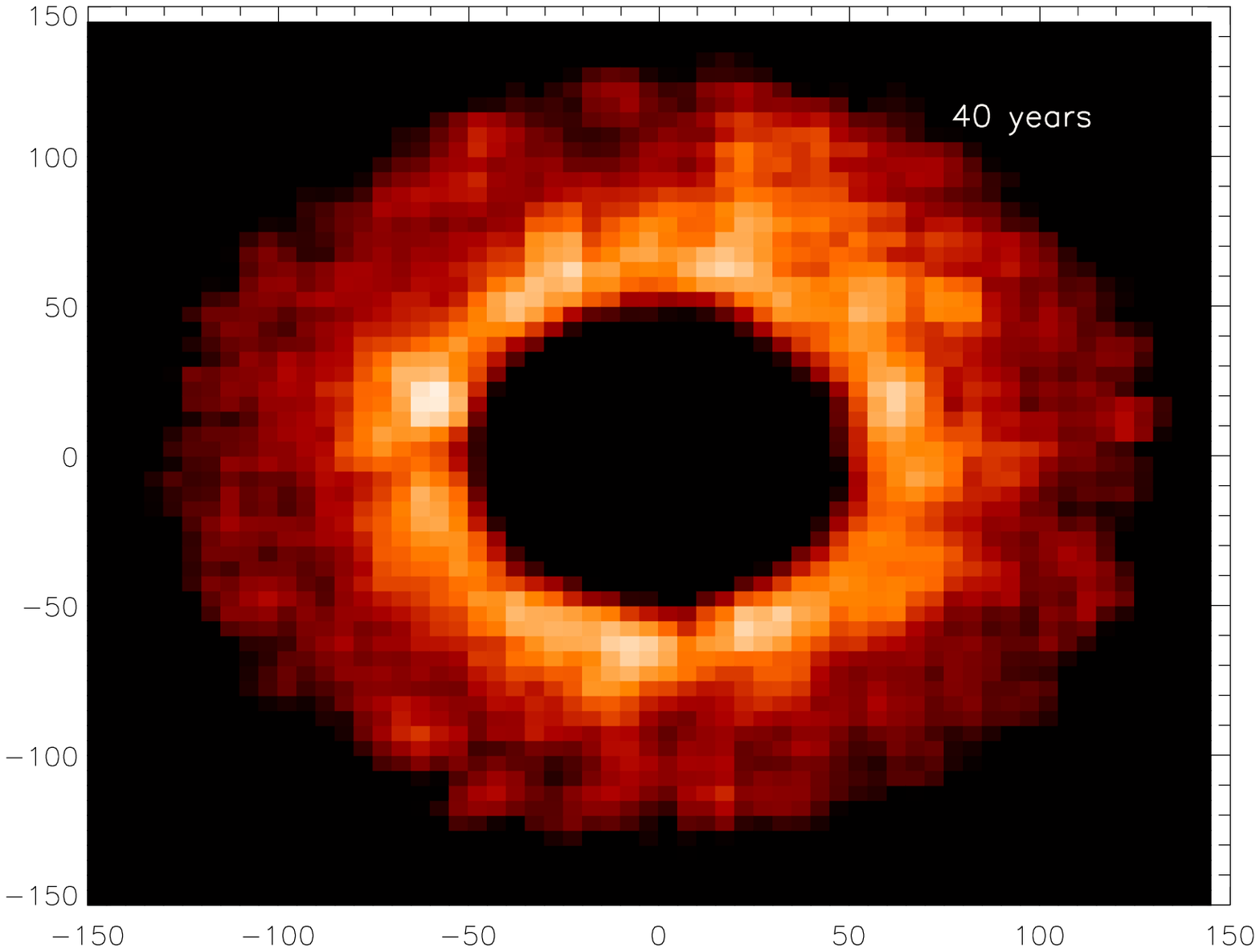}
\includegraphics[scale=0.5]{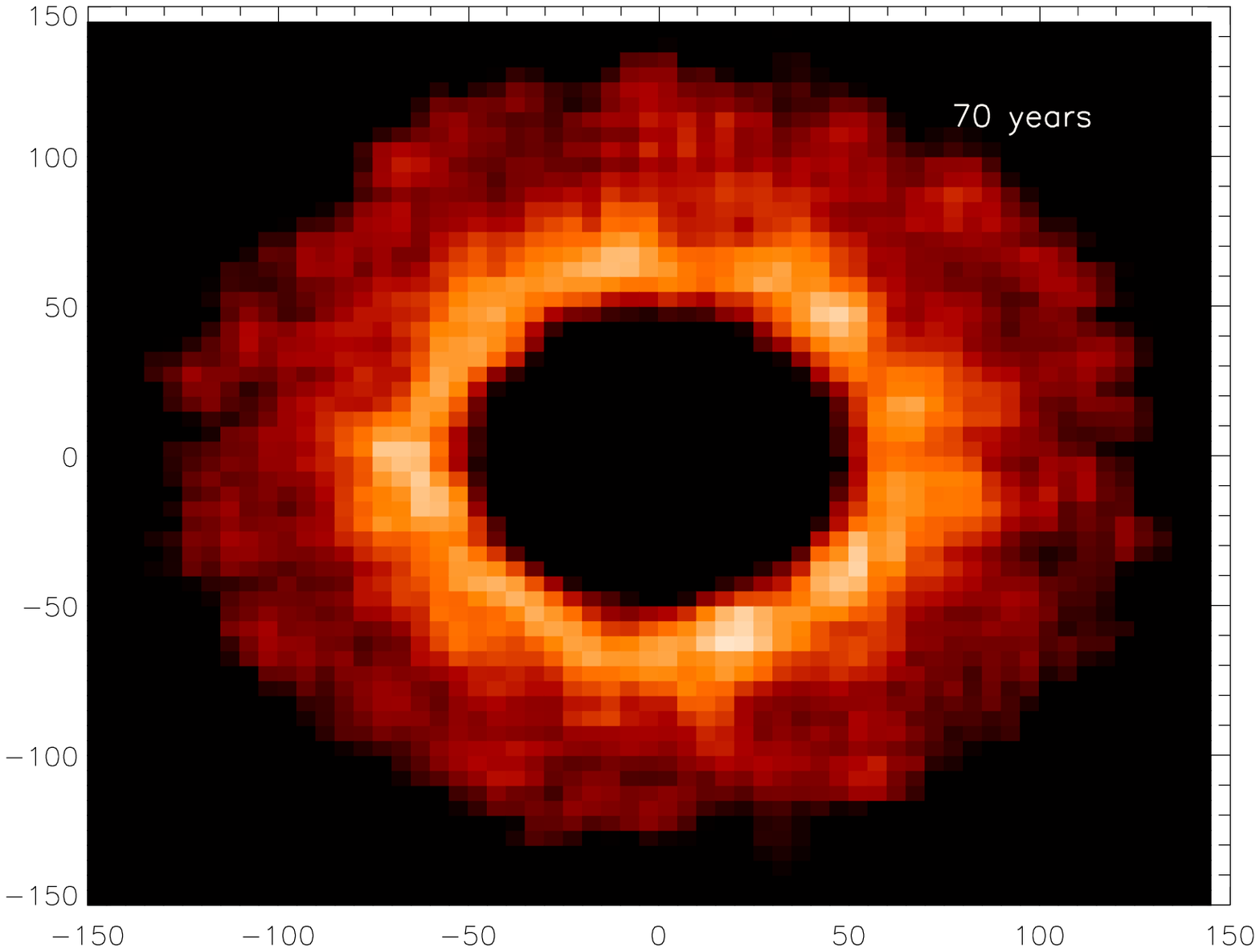}
}
\caption[]{Normalized synthetic images, in scattered light, showing the evolution of an avalanche in a cold debris disc (extending from 50 to 120\,au) of average optical depth $\tau_{\perp}$$\sim$2$\times 10^{-3}$, for an initial breakup, at $t$=0, releasing $5\times$10$^{22}$g of dust at $10\,$au from the star.  }
\label{avalnom}
\end{figure*}

\begin{figure}
\includegraphics[scale=0.5]{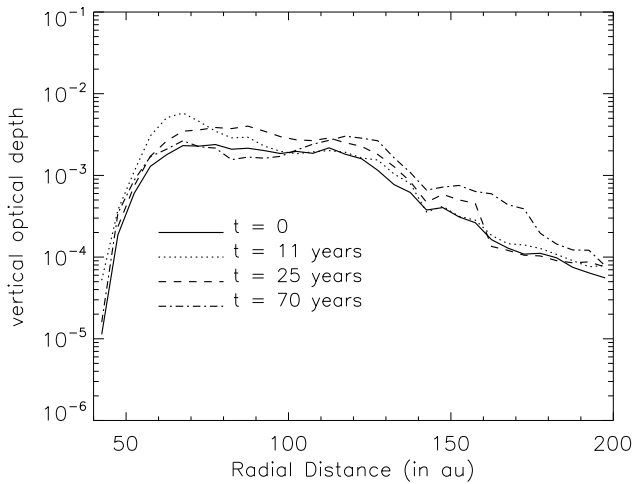}
\caption[]{Radial cuts of the vertical optical depth, passing by the avalanche's peak luminosity location, at 4 different epochs of the avalanche's evolution. The profile at t = 0 serves as a reference profile to which the avalanche-induced dust-excess can be compared.}
\label{cut}
\end{figure}

\begin{figure}
\includegraphics[scale=0.5]{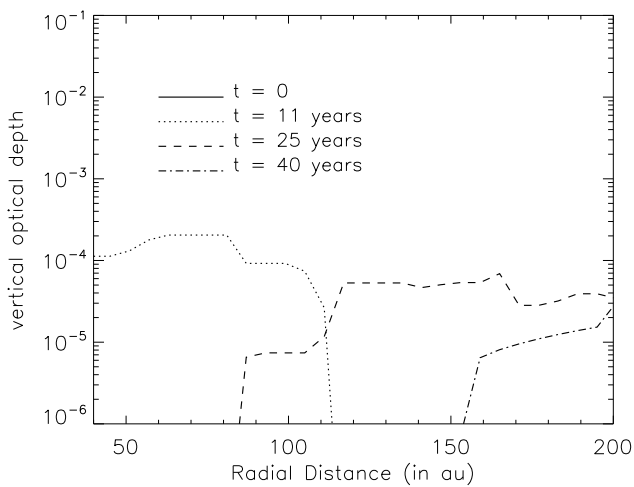}
\caption[]{Same as Fig.\ref{cut}, but for a test simulation with no outer disc, that is, showing only the dynamical outward motion of the grains released by the initial breakup event at 10\,au. It can be clearly seen that the front of this outward propagation is, after only 40\,years, already well beyond 200\,au (we show the radial profiles up to $t=40$\,years instead of the 70\,years of Fig.\ref{cut}, because there are no high-$\beta$ initial grains left in the <200au region after 70 years).}
\label{cnc}
\end{figure}

\begin{figure}
\includegraphics[scale=0.5]{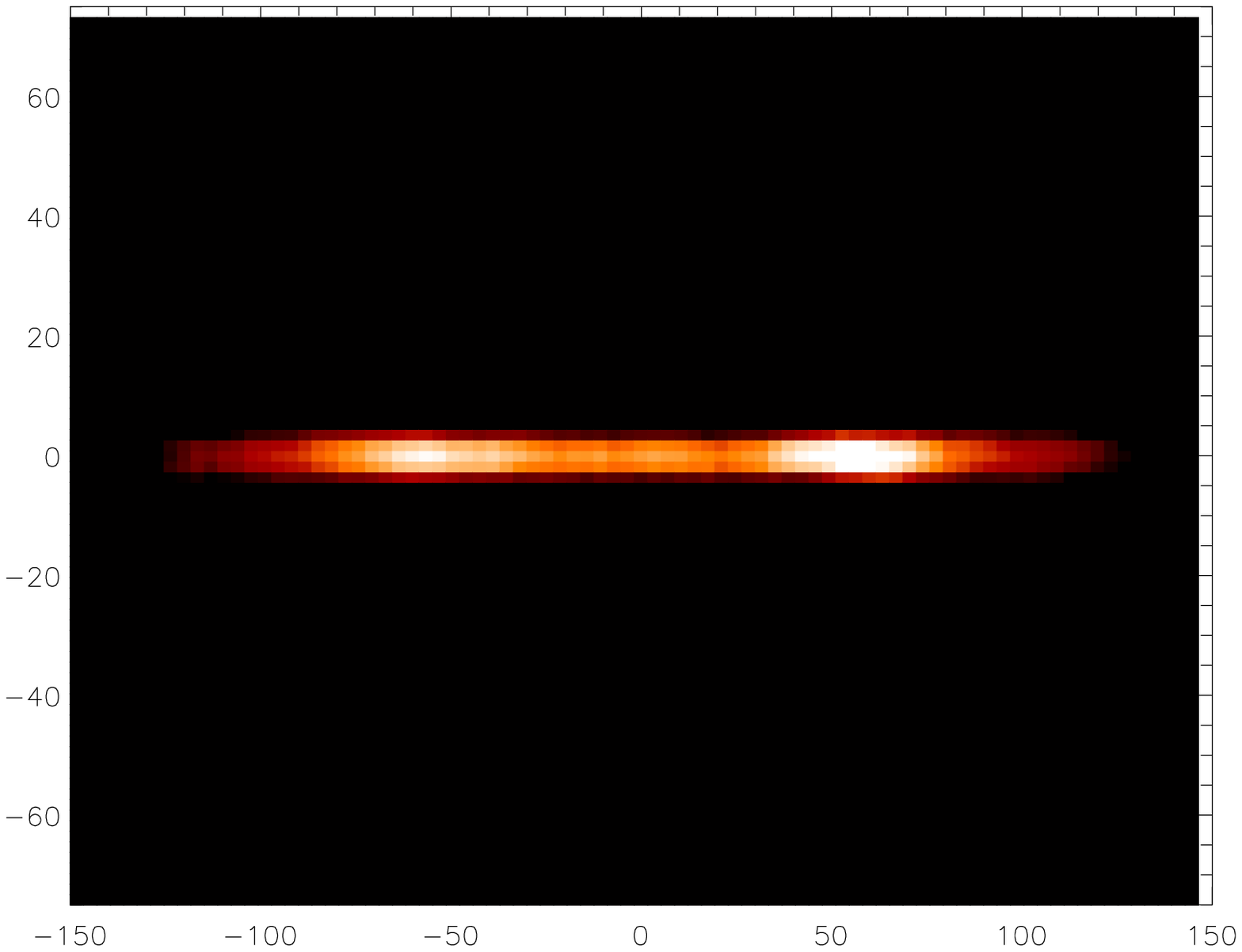}
\caption[]{Cold disc case: Scattered light synthetic image, at the time of the maximum brightness of the avalanche (t=11\,years), for the same case as in Fig.\ref{avalnom} but for a system viewed edge-on.}
\label{edge}
\end{figure}

\subsection{"Cold outer disc'' case}

\subsubsection{Reference GAT07 configuration} \label{gat07}

\begin{figure}
\includegraphics[scale=0.5]{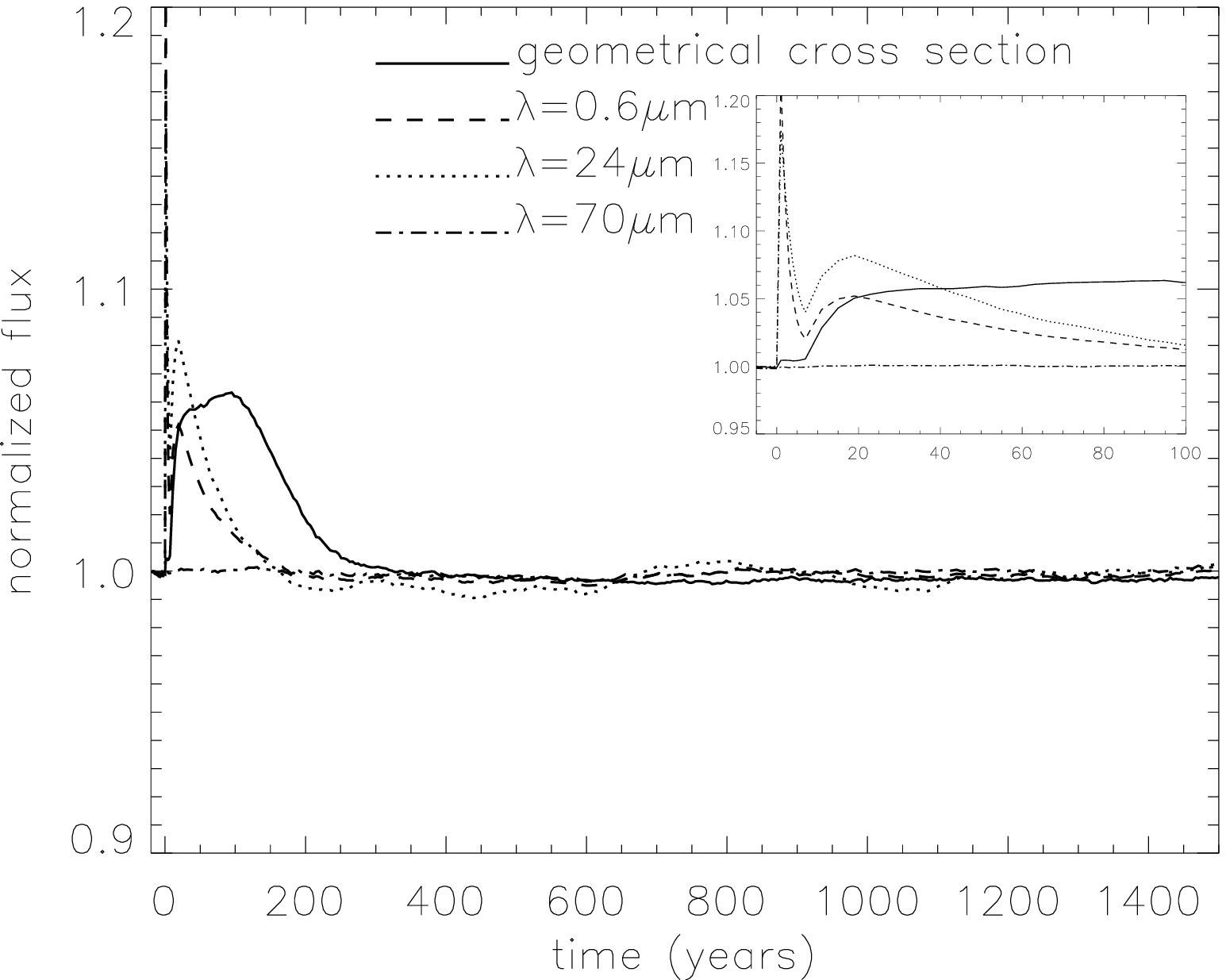}
\caption[]{Cold disc case: disc-integrated luminosity as a function of time, at 3 different wavelengths, as well as for the total geometrical cross section. All fluxes are normalized to their value before the onset of the avalanche. The inset figure is a zoom-in on the first 100 years of the simulation. The brief initial spike corresponds to the luminosity of the dust that is released at the moment of the breakup. }
\label{fluxcoldgrig}
\end{figure}

\begin{figure}
\includegraphics[scale=0.5]{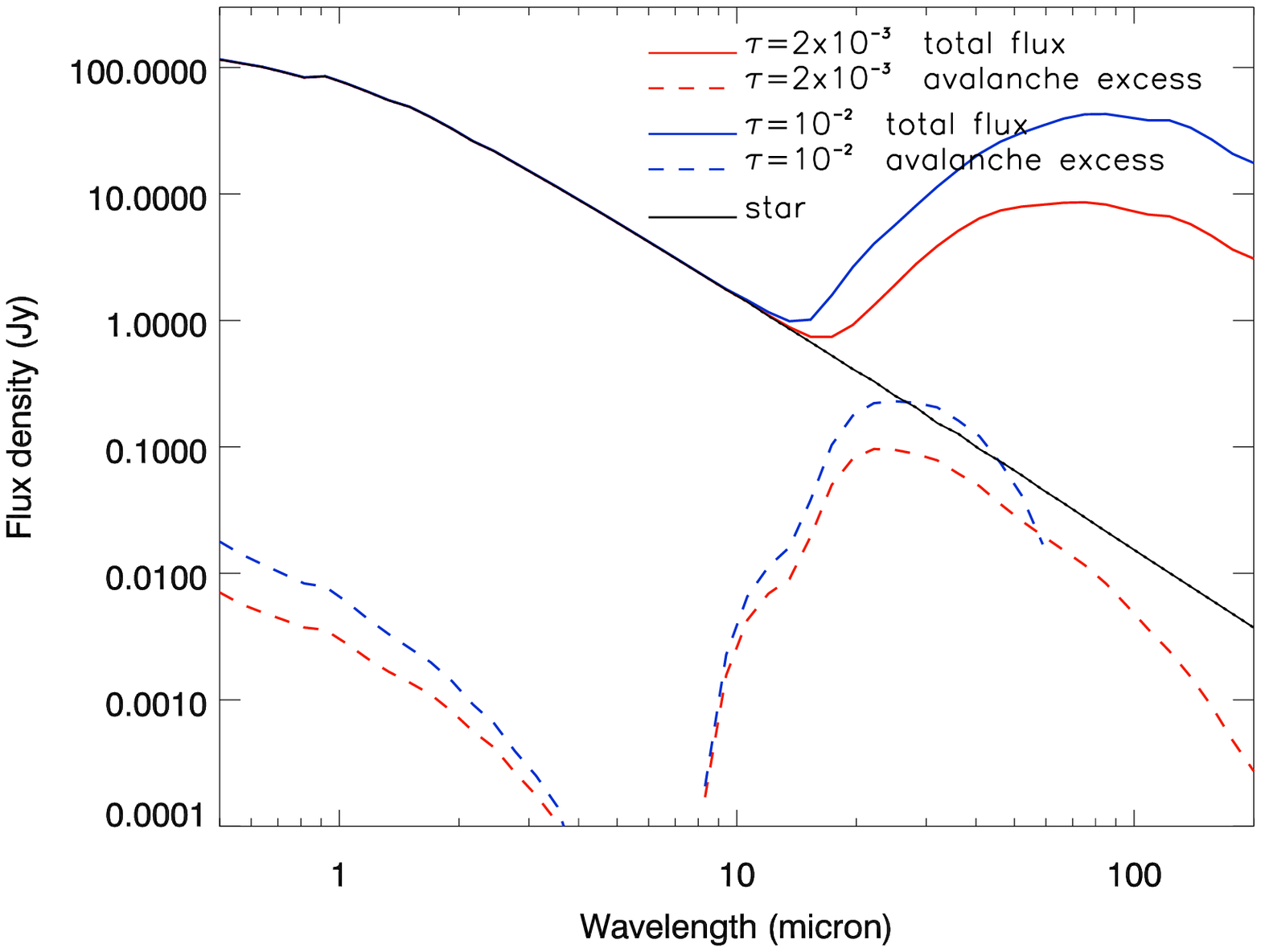}
\caption[]{Cold disc case: system-integrated SED (using the GRaTer package for a system at a distance of 30pc), at the peak of the avalanche luminosity, for the reference GAT07 $\tau_{\perp}\sim$2$\times 10^{-3}$  disc and for a denser $\tau_{\perp}\sim$10$^{-2}$ disc. The dashed lines give the luminosity excess due to the avalanche (computed by subtracting the SED of a reference avalanche-free case). The cutoff of the dashed-blue line beyond $50\mu$m reflects the fact that, for this dense disc case, the avalanche-system is slightly less luminous, at these long wavelengths, than an avalanche-free counterpart. This is because, beyond $50\mu$m, the flux is mostly dominated by large grains, which are slightly eroded by the passage of the avalanche (see discussion in Sec.\ref{drop})}
\label{SEDcold}
\end{figure}

We first consider the same set-up as GAT07, i.e., a "cold disc'' case with an average vertical optical depth $\tau_{\perp}\sim$2$\times 10^{-3}$ (i.e. $\tau_{r} \sim$2$\times 10^{-2}$) and $m_{\rm{init}}=10^{20}$g of dust released by the breakup. We confirm that, in this case, the effect of an avalanche is negligible. In fact, at all considered wavelengths, the pattern that it potentially leaves in the outer disc is below the natural noisiness of the disc as simulated by LIDT-DD. If we follow GAT07 and take as a simplified threshold for avalanche observability that a local contrast of $\sim100$\% with the background disc can be measured at some location in scattered light images, we find that, to meet this criteria, the released amount of dust has to be very large, $m_{\rm{init}}\sim$5$\times10^{22}$g, which is equivalent to the mass of a $\sim$150\,km object. 

Fig.~\ref{avalnom} shows synthetic head-on images of the system in scattered light, for an unfolding avalanche in this massive $m_{\rm{init}}\sim$5$\times10^{22}$g breakup case. It is qualitatively similar to what was obtained by GAT07, with a tail of high-$\beta$ grains escaping on a spiral-like trajectory from the breakup location, and then penetrating the outer disc where they produce a collisional chain-reaction (the avalanche) that leaves a bright outward propagating trail. The main difference with GAT07 is that, while this earlier study, by artificially separating avalanche grains from field particles, was able to visualize the unfolding avalanche regardless of its actual strength, we have a more realistic estimate of an avalanche's actual magnitude, for which it only becomes detectable, on synthetic images, for the much larger amount of released dust considered here. 
Our more realistic approach also explains why the avalanche appears to fade in the outer parts of the disc (Figs.\ref{avalnom}d,e,f). This is not because the avalanche stops propagating or becomes weaker in terms of the relative excess of small grains it can locally create, but because these outer regions are fainter, in scattered light, due to the geometrical flux dilution. To illustrate this point, we plot in Fig.\ref{cut} the evolution of the radial profile of the disc's vertical optical depth at the approximate longitude of the avalanche \footnote{Note that these different cuts are not along the same axis because the avalanche's longitude evolves with time}. The outward propagation of the avalanche is here much more apparent, even in the outer regions. As a matter of fact, the $\tau_{\perp}$ \emph{contrast} between the outward-moving peak of the avalanche-induced excess and the background disc is relatively constant with time, of the order of a factor 2, and does not decrease as the avalanche propagates. What decreases is the luminosity of the regions reached by the outward propagating avalanche. Note also that the regions reached by the avalanche continue to display an excess  $\tau_{\perp}\,$ after the passage of the avalanche front "wave".
Another important result is that, even if the outward propagation of the avalanche is relatively fast, it is, however, much slower than the purely dynamical outward movement of the initial grains through the disc (Fig.\ref{cnc}). This means that the avalanche does not only consist of first generation fragments created by collisions with the initial "trigger" grains sweeping through the disc, because in this case its outward propagation would simply follow that of these trigger grains. Its much slower propagation confirms that it is in fact a chain reaction involving several successive generations of fragments.

If the system was to be seen edge-on (Fig.\ref{edge}), the maximum avalanche-induced luminosity contrast drops to $\sim$30\,\%, about one third of its value in the head-on images.

If we now consider the disc-integrated photometry (Fig.~\ref{fluxcoldgrig}), we see that, even for this very massive breakup event, the peak avalanche-induced luminosity excess is relatively limited, comprised between 0.2\% (at $\lambda$=70$\mu$m) and 8\% (at $\lambda$=24$\mu$m) in thermal emission, and is $\sim$5\% in scattered light ($\lambda$=0.6$\mu$m). This also means that the luminosity excess is at its weakest at the wavelength where this cold disc is the brightest, i.e., around 70$\mu$m, as appears clearly when plotting the system's Spectral Energy Distribution (SED) (red curve in Fig.~\ref{SEDcold}), where we see that the avalanche-induced luminosity rapidly drops beyond $\lambda$$\sim$40$\mu$m. This is due to the fact that the dust particles produced by the avalanche are very small grains, mostly $\leq 1\mu$m, which, even if they contribute to most of the avalanche-induced excess in terms of geometrical cross section, are very poor emitter at long wavelengths. At $24\mu$m, however, this decreased emissivity is compensated by the fact that these tiny grains are warmer than the larger grains of the main disc and that we are here in the Wien domain of the Planck function, for which a temperature increase results in an exponential increase of the $B_{\lambda}(\lambda,T)$ parameter.

In addition, it is important to note that, for this reference $\tau_{\perp}\sim$2$\times 10^{-3}$ case, the breakup event \emph{itself} creates a luminosity excess which exceeds, at all wavelengths, that of the subsequent avalanche (see inset in Fig.~\ref{fluxcoldgrig}). In other words, from a photometric point of view, the avalanche is not able to amplify the signal produced in the immediate aftermath of the breakup. This seems to be in disagreement with GAT07, who obtained an "amplification" factor $F$$\sim$200 for a similar set-up. This difference is mostly due to the fact that $F$ measures geometrical cross sections and \emph{not} actual luminosities. As such it does not account for the fact that the initial breakup occurs closer to the star, where both the scattered light (because of the stellar flux dilution in $1/r^{2}$) and thermal (because dust closer to the star is warmer and thus more luminous, at least in the 24-70$\mu$m range) emission of a given dust grain are higher than further out in the disc. When considering the disc-integrated total geometrical cross section, we retrieve GAT07's results, which is that the avalanche does indeed have an amplification effect when compared to the initially released dust population (full line in Fig.~\ref{fluxcoldgrig}).

What the avalanche does, however, is to extend the duration of the luminosity excess period. Indeed, while the luminosity excess created by the breakup itself quickly drops to a much lower level, the subsequent avalanche-induced excess lasts for a much longer time, of the order of $\sim$100\,years, roughly a third of the dynamical timescale at the innermost part of the main disc. This duration is, however, shorter than what was found by GAT07 (see Fig.3 of that paper), a difference that is once again due to the fact that GAT07 were considering the system's cross section and not its photometry. Indeed, the cross section amplification happening in the outer parts of the disc, which strongly contributes to the extended period of excess when plotting this parameter (see also Fig.\ref{cut}), concerns grains whose contribution to the system's IR-photometry (be it in thermal emission or scattered light) is much less important.

\subsubsection{Denser outer disc} \label{grigdens}

\begin{figure}
\includegraphics[scale=0.5]{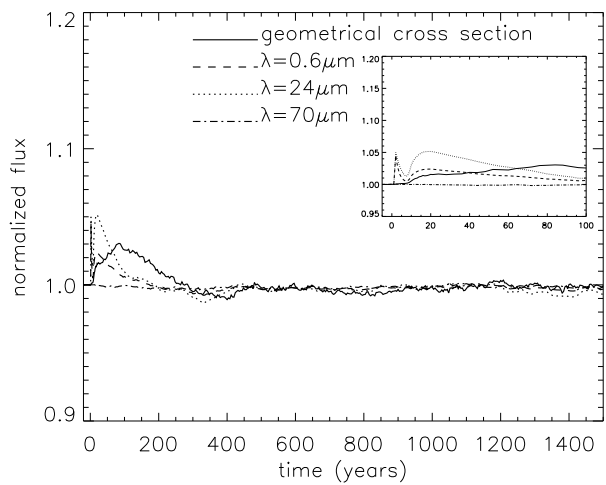}
\caption[]{Same as Fig.~\ref{fluxcoldgrig}, but for a denser main disc with $\tau_{\perp}\sim$10$^{-2}$, all other parameters (released dust mass, disc location and width) being the same. The initial luminosity flash due to the breakup event itself appears less pronounced here because all luminosities are rescaled to the main disc's brightness, which is here higher than for our previous reference case.}
\label{fluxcoldbest}
\end{figure}

We now increase the density in the outer disc by a factor $\sim$5, i.e., to values comparable to those measured for the "extreme'' discs, with $\tau_{\perp}$ of the order of $10^{-2}$ (i.e., $\tau_{r} \sim$10$^{-1}$). To allow easy comparison, we consider the same massive initial dust release ($m_{\rm{init}}\sim$5$\times10^{22}$g) that was the threshold for the avalanche to become visible in the previous reference case. We see that, this time, the disc-integrated luminosity of the avalanche exceeds that of the initial breakup at $\lambda$=24$\mu$m (Fig.~\ref{fluxcoldbest}). 
However, this luminosity amplification over the initial breakup remains relatively limited, barely reaching a factor $\sim$1.1, and being still below $1$ at all other wavelengths (see inset in Fig.\ref{fluxcoldbest}).
In fact, if we consider the \emph{relative} luminosity ratio between the avalanche and the main disc, we see that this ratio is actually slightly lower, for the same amount of post-breakup released dust, than in the previous case of a more tenuous disc (Fig.~\ref{fluxcoldbest}). Likewise, on synthetic 2D-images (not shown here), the maximum local contrast in scattered light with the main disc is now only $\sim$60\,\% instead of the $\sim$100\,\% obtained for the $\tau_{\perp}=2\times10^{-3}$ disc. 

\begin{figure}
\includegraphics[scale=0.5]{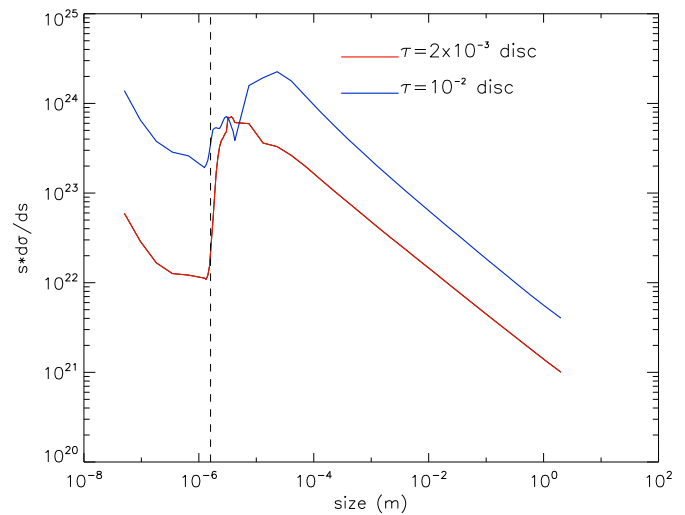}
\caption[]{Cold disc case: Disc-integrated geometrical cross section per logarithmic size bin, for the disc at collisional steady-state just \textit{before} the onset of the avalanche. The vertical dashed line marks the $\beta=0.5$ limit for unbound particles (if produced from progenitors on circular orbits)}
\label{psdcold}
\end{figure}

This seems to sharply contradict the results of GAT07, who estimated that an avalanche's amplitude should exponentially vary with the main disc's density, and should increase by almost 2 orders of magnitude when increasing $\tau_{r}$ by a factor 4 (see Fig.14 of that paper). 
This fundamental difference with GAT07 is due to one simplification that was made in that paper, which is that their main disc was devoid of unbound grains. While this approximation might be justified for systems with low $\tau_{\perp}$, it becomes much more questionable for denser discs. As an illustration, we see that, for the present $\tau_{\perp}\sim$10$^{-2}$ case, unbound grains are not at all negligible in the main outer disc, and even marginally dominate its total geometrical cross section (Fig.~\ref{psdcold}). As a consequence, the \emph{additional} influx of unbound grains coming from the planetesimal breakup is much less "felt" by the main disc. Granted, these avalanche unbound grains have a much higher destructive power because of their higher outward velocities, but they are not numerous enough, even for a massive dust release of $m_{\rm{init}}\sim$5$\times10^{22}$g, to create an important difference. An additional effect that weakens the avalanche's amplitude is that, because of the presence of these unbound grains in the main disc at steady-state, this disc is depleted of small bound grains in the $\sim$1--10$\mu$m range (Fig.~\ref{psdwarm}), which correspond to the grains that are mostly eroded by the unbound grains of the avalanche and thus make up the reservoir to fuel the avalanche's evolution.

\subsection{"Warm disc'' case} \label{warm}

As explained in the Sect.\ref{model}, this case is scaled down, in stellar distance, by a factor 10 as compared to the previous case.
In order to have discs which fractional luminosities equivalent to those in the cold disc case, we scale down the main disc's mass by a factor of 100. This means that, in terms of optical depths, we consider the same 2 cases as before, i.e., a reference main disc with $\tau_{\perp}$$\sim$2$\times 10^{-3}$  and a bright disc case with $\tau_{\perp}$ of the order of 0.01.

\subsubsection{Reference $\tau_{\perp}$$\sim$2$\times 10^{-3}$ disc}

\begin{figure}
\includegraphics[scale=0.5]{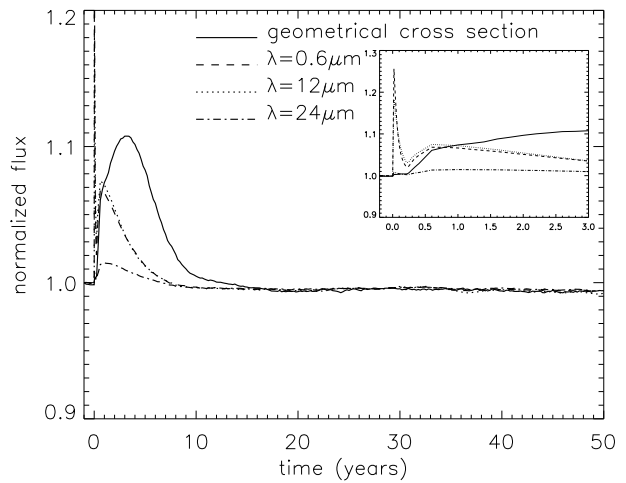}
\caption[]{Same as Fig.~\ref{fluxcoldgrig}, but for the "warm disc" case (outer main disc between 5 and 12au with an optical depth $\tau_{\perp}$$\sim$2$\times10^{-3}$), and a mass  $m_{\rm{init}}$$\sim$10$^{20}$g  of dust released at 1ua. }
\label{fluxinner}
\end{figure}

\begin{figure}
\includegraphics[scale=0.5]{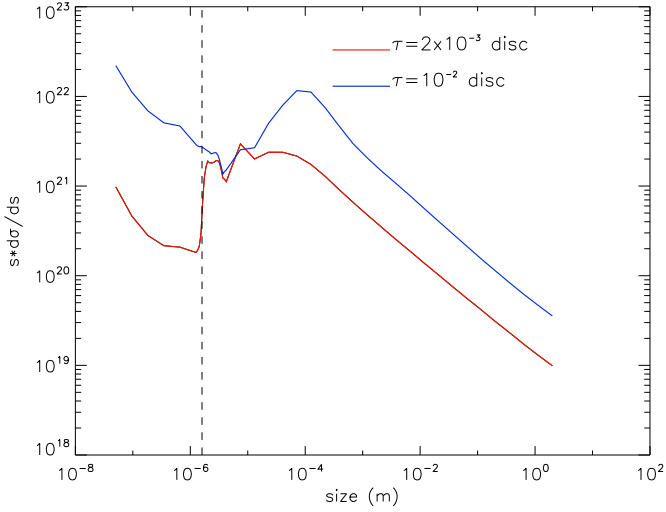}
\caption[]{Warm disc case: Disc-integrated geometrical cross section per logarithmic size bin, for the disc at collisional steady-state just before the onset of the avalanche.}
\label{psdwarm}
\end{figure}

Not surprisingly, applying the same criteria of a local surface scattered-light luminosity contrast of $\sim$100\% with the background disc, we find that the avalanche becomes significant for a smaller released dust mass than in the cold disc case. We find, however, that this 100\% contrast threshold is reached for $m_{\rm{init}}$$\sim$10$^{20}$g, which is a factor $\sim5$ less than what would have been found by scaling down $m_{\rm{init}}$ by the same factor 100 as the disc's total mass. This can be explained by the fact that impact velocities are much higher than for the cold disc case, because we are here 10 times closer to the star, so that the destructive power of avalanche grains, and the number of debris they produce, are much higher as they sweep through the disc. For an equivalent set-up, avalanches are thus more powerful in warm discs. They would in fact be even more powerful if their strength was not reduced by the large population of unbound grains already present in the outer disc \emph{before} the avalanche's passage (Fig.\ref{psdwarm}). The avalanche-dampening effect due to these "native" unbound grains (in particular the depletion of avalanche-fueling target grains in the $1-10\mu$m range) had also been observed for the cold disc case, but only for a dense disc with $\tau_{\perp}$$\sim$10$^{-2}$ (see Sec.\ref{grigdens}), whereas it is here already noticeable for $\tau_{\perp}$$\sim$2$\times10^{-3}$. This important population of unbound grains at lower $\tau_{\perp}$ is the direct consequence of more violent collisions (because of higher $v_{col}$) in the main disc itself.

\begin{figure}
\includegraphics[scale=0.5]{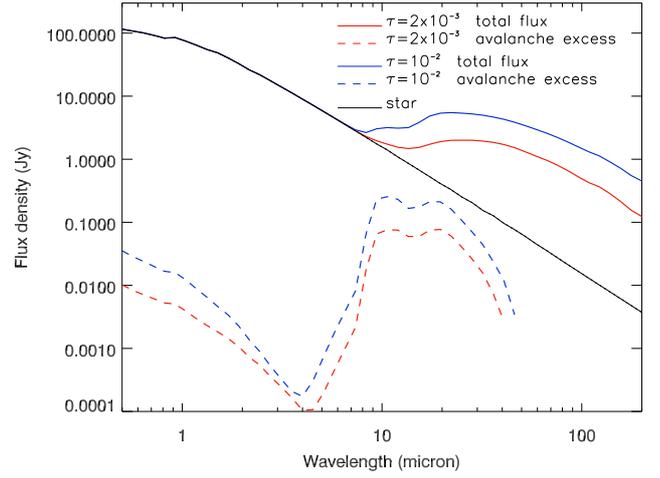}
\caption[]{Warm disc case: system-integrated SEDs, for two different outer disc densities, at the time the avalanche reaches its peak luminosity.}
\label{SEDwarm}
\end{figure}

Note, however, that the avalanche's "amplification" power, in terms of its peak luminosity as compared to that of the initially released dust is of the same order as for the $\tau_{\perp}$$\sim$2$\times10^{-3}$ cold disc case, with the initial post-breakup "flash" being here also brighter, at all considered wavelengths, than the subsequent avalanche (see inset in Fig.\ref{fluxinner}). 
The main effect of the avalanche is here again to produce a luminosity excess that lasts longer than the very transient initial peak. Its duration is of the order of 3-4 years, which is, logically, roughly the $\sim$100\,years timescale of the cold disc scaled down by a factor (100\,au/10\,au)$^{1.5}$.
The maximum relative photometric excess due to the avalanche is of the order of 7--8\% and peaks in the $\sim$10--15$\mu$m wavelength domain (Fig.\ref{SEDwarm}). Interestingly, and contrary to the cold disc case, this wavelength domain is also roughly that where the disc itself is the brightest.

In any case, the onset of an observable avalanche still requires the initial release of a very large amount of dust mass, $m_{\rm{init}}$$\sim$10$^{20}$g being equivalent to the the mass of a $\sim$20\,km sized object, which, combined to the fact that the avalanche is more short-lived than in the cold disc case, might not render this 5-12au case as avalanche-friendly as it appears (see Sec.\ref{discu}) 

\subsubsection{Denser disc}

\begin{figure}
\includegraphics[scale=0.5]{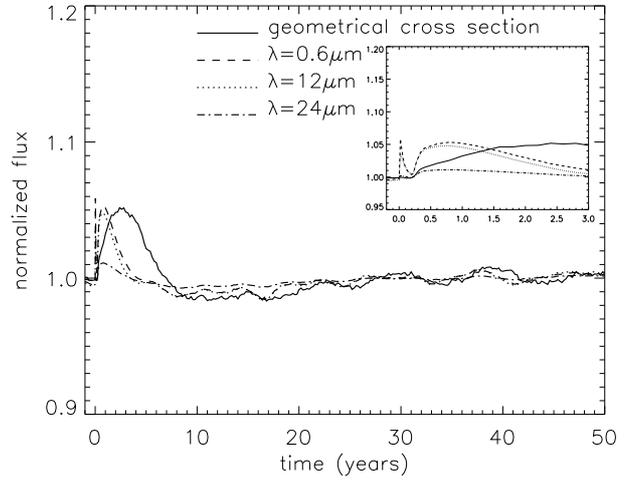}
\caption[]{Same as Fig.~\ref{fluxinner}, but for a denser disc with optical depth $\tau_{\perp}\sim$10$^{-2}$).}
\label{fluxinnerd}
\end{figure}

As for the cold disc case, increasing the main disc optical depth to $\tau_{\perp}$$\sim$10$^{-2}$ does not lead to the much stronger avalanches expected from using the GAT07 empirical relations. Granted, the avalanche does get somehow stronger in absolute magnitude than for a $\tau_{\perp} $$\sim$2$\times10^{-3}$ disc, with its peak luminosity now slightly exceeding that of the initial post-breakup flash (Fig.\ref{fluxinnerd}). However, its \emph{relative} luminosity with respect to the main disc is lower. As for the cold disc case, this unexpected result is due to the crucial role played by "native" unbound grains already present in the main disc. As Fig.\ref{psdwarm} indeed clearly shows, in a 5-12au disc with $\tau_{\perp}$$\sim$10$^{-2}$ at steady state, the geometrical section is dominated by $\beta$>0.5 grains, even more so than in our cold disc case (Fig.\ref{psdcold}). As a consequence, both effects due to these native unbound grains (denser backdrop onto which avalanche unbound grains are added and depletion of avalanche-fueling targets in the $1-10\mu$m range) are more pronounced here and concur to weaken the avalanche.

\section{Discussion} \label{discu}

\subsection{dependence on disc density} \label{densdep}

One crucial difference with Grigorieva et al.(2007) is that we do not get significantly stronger avalanches when increasing the density of the disc within which they propagate, whereas GAT07 found that avalanche strength exponentially increases with $\tau_{r}$ (and thus with $\tau_{\perp}$ for most non "exotic" disc configurations, see Equ.\ref{tau}). As explained in Sec.~\ref{grigdens}, this is mostly because dense discs ($\tau_{\perp}$$\sim$10$^{-2}$) already contain large amounts of unbound grains \emph{before} the passage of the avalanche, a fact that was not accounted for in GAT07. These native unbound grains weaken the avalanche in two ways, firstly by creating a dense background that reduces the relative excess density due to avalanche-induced unbound grains, and secondly by preemptively eroding the population of small bound grains from the main disc (in the $\sim$1--10$\mu$m range) which makes up most of the mass reservoir for fueling the avalanche propagation. This means that the empirical relation derived by GAT07 for estimating the strength of an avalanche (or more exactly its cross section "amplification factor" $F$) as a function of disc density cannot be used in realistic systems (even when overlooking the fact that $F$ is a very imperfect and ambiguous measure of an avalanche's amplitude).

As a consequence, avalanches do not get easier to observe in discs denser than $\tau_{\perp}$$\sim$2$\times 10^{-3}$. This means that, contrary to what could have been expected, very bright "extreme" discs are probably not such a favourable ground for avalanche propagation (see Sec.\ref{extreme})
Conversely, additional test simulations show that avalanche amplitude quickly drops for discs more \emph{tenuous} than our  $\tau_{\perp}$$\sim$2$\times$10$^{-3}$ reference case. This is an expected result, as the breakup-released unbound grains will then basically cross a faint disc without encountering any resistance, making this case very similar in essence to that of a large breakup in a dust-free system \citep[see, for instance][]{kral15}.

This makes discs with optical depths of a few $10^{-3}$ as probably the optimum cases for maximizing the contrast between an avalanche and the disc it propagates through.

\subsection{photometric signature and wavelength dependence}

A new result from the present study is that we show that, for all considered cases, avalanches have a relatively limited effect on disc-integrated photometry. For a cold disc, even the avalanche triggered by the massive release of $5\times$10$^{22}$g of dust does not create a luminosity excess of more than $\sim$10\%. Moreover, at $\lambda$>50$\mu$m wavelengths where the disc is the brightest, the avalanche-induced excess drops to almost zero because of the poor emissivity of the tiny avalanche grains in the far IR. 
For warm discs at $\sim10$\,au, the situation becomes more favourable, with the photometric avalanche-induced excess peaking in the $\lambda$$\sim$10--$20\mu$m domain where the system is the brightest (Fig.\ref{SEDwarm}). However, this luminosity increase is, here again, limited to less than $\sim$10\%.

Such 5--10\% photometric variations in the 10--20$\mu$m wavelength domain could, however, be within the reach of mid-IR observing facilities such as the Wide-field Infrared Survey Explorer (WISE), which has a relative photometric accuracy of less than 10\% at these wavelengths \citep{melis2012}, or the Spitzer MIPS instrument, having an accuracy of $\sim$2\% at 24$\mu$m \citep{rieke08}.

Of course, these luminosity excesses could become more prominent when considering even larger masses of initially released dust $m_{\rm{init}}$, because avalanche amplitudes scale linearly with $m_{\rm{init}}$ (GAT07). There is, however, a trade-off, as increasing the mass of the initial breakup object strongly decreases the likelihood for such a breakup to occur (see next section). Moreover, above a given size the breakup object becomes large enough to trigger, in addition to the initial short-lived luminosity "flash" mainly due to the unbound grains it releases, a collisional cascade amongst its own debris that is brighter and much more long lived than the avalanche. Such a post-breakup collisional cascade might thus be much easier to detect \citep[see, for instance, the simulations of][investigating the aftermath of the destruction of a Ceres-like planetesimal]{kral15} than the avalanche.

\subsection{duration and likelihood} \label{likely}

For a large enough initial dust release, an avalanche is able to leave an observable signature (both in photometry and on resolved images) on a timescale $t_{\rm{aval}}$ that is typically a fraction of an orbital period in the main outer disc. We find $t_{\rm{aval-cold}}$$\sim$100\,years for the cold disc and $t_{\rm{aval-warm}}$$\sim3\,$years for the warm disc cases, respectively. Not surprisingly, the ratio between these 2 timescales is close to the one between dynamical timescales for these two configurations, i.e., $10^{1.5}$. As a consequence, we estimate that the typical duration for an avalanche follows the empirical law $t_{\rm{aval}}$$\sim$100$\,(r_{\rm{disc-in}}/50\rm{au})^{1.5}$, or $t_{\rm{aval}}$$\sim$0.3$\,t_{\rm{orb}-in}$, where $r_{\rm{disc-in}}$ is the location of the inner edge of the main outer disc and $t_{\rm{orb}-in}$ the orbital period at this location.

Note that $t_{\rm{aval}}$ exceeds the time it takes for small unbound grains with high $\beta$, produced at the breakup's location, to directly cross the main disc. Using the asymptotic limit for the radial velocity of such a grain ($v_{\rm lim}=v_{\rm kep} (2\beta-1)^{0.5}$, see \cite{seze17}), we find that, for the cold disc case, a grain produced at 10\,au with $\beta=5$ crosses the 50 to 120au radial extension of the disc in less than 10 years, 10 times less than the photometric duration of an avalanche (see Figs.~\ref{fluxcoldgrig} and \ref{fluxcoldbest}) \footnote{this $\sim$1/10 ratio in timescales holds also for the warm disc case}. This point is clearly illustrated by the comparison of Figs.\ref{cut} and \ref{cnc}.

However, $t_{\rm{aval}}$ remains extremely short when compared to the typical age $t_{age}$ of even the youngest debris discs. This means that, in order to be observed, an avalanche-producing breakup event needs to occur on a typical timescale $t_{\rm{break}}$ that does not greatly exceed $t_{\rm{aval}}$. The probability of witnessing an avalanche is then, for a given system, of the order of $t_{\rm{aval}}/t_{\rm{break}}$. The problem is that  $t_{\rm{break}}$ is expected to sharply decrease with object sizes and that our simulations have shown that the threshold for avalanche detectability is only reached for the breakup of very large bodies. If we follow GAT07 and make the relatively optimistic hypothesis that 10\% of the shattered object is converted into $s\leq 1\,$cm dust, then we need to shatter a $\sim$300\,km object around 10\,au to trigger an observable avalanche in a cold disc, and a $\sim$40\,km object around 1\,au for a warm disc. 
Estimating the shattering frequency of such objects is a complex task, as it might depend on several poorly constrained, or strongly varying from system to system, parameters, such as the size distribution of large objects in the system, their average dynamical excitation, etc... For the sake of simplicity, we consider the asteroid belt as a proxy for a typical disc of kilometre-sized bodies in the 1-10\,au region \footnote{We are aware that it is an oversimplification to consider the asteroid belt as a proxy for belts hosting avalanche-triggering events for both our "cold" and "warm" disc cases. We are hereby overlooking that the avalanche-releasing breakup occurs at $\sim$10\,au in the former and at $\sim$1\,au in the latter, and that the conditions for large planetesimal breakups might be different at these two locations. However, we will not enter into this complex issue here and keep the asteroid belt as a satisfying first-order middle ground between these two locations}. 
If we take the estimates for the interval between disruptions in the main belt as a function of asteroid sizes derived by \citet{bott05}, we get $t_{\rm{break}(s>\rm{300km})}\sim$3$\times10^{9}$years and $t_{\rm{break}(s>\rm{40km})}\sim$3$\times10^{7}$years, respectively. These values exceed by several orders of magnitudes our $t_{\rm{aval}}$ estimates (100\,years and 3\,years, respectively), making avalanche-inducing breakups very unlikely events in asteroid-like belts. 

However, the asteroid belt is a relatively tenuous and mass-depleted disc, and breakup rates could be higher in denser systems. All other parameters (distance to the central star, profile of the size distribution, impact speeds) being equal, the rate of catastrophic breakups should scale as $(M_{disc}/M_{ast})^{2}$, where $M_{ast}$ is the total mass of the asteroid belt, implying that a disc $\sim$3$\times10^{3}$ more massive than the asteroid belt is needed to reach $t_{\rm{break}(s>\rm{40km})} = t_{\rm{aval-warm}}$, with $(M_{disc}/M_{ast})\sim$5$\times10^{3}$ being necessary to get $t_{\rm{break}(s>\rm{300km})} = t_{\rm{aval-cold}}$ 

Note that the cold and warm disc estimates for $(M_{disc}/M_{ast})$ are relatively close, because the smaller size of the object that has to be broken in the warm disc case is compensated by the fact that an avalanche is then more short lived and has to occur more frequently than for the cold disc case. Considering that the current asteroid belt has a fractional luminosity of $f\sim$10$^{-7}$ \citep{derm02}, it would imply that such an "enhanced" belt would have a fractional luminosity of a few $10^{-4}$, which is within the range of observed fractional luminosities of warm discs \citep[][and references therein]{2014ApJS..211...25C,2017arXiv170302540K} 

In this case, an ideal configuration for an "avalanche-friendly" system could be a two-belt structure with a relatively bright ($f\gtrsim$10$^{-4}$) asteroid belt around 1\,au (warm disc case) or 10\,au (cold disc case), and an extended disc further away from the star. Such a configuration would be very similar to the ones investigated by \citet{geil17}, who identified several double-belt systems fitting this description (see Fig.~2 of that paper).%, at least for the inner-belt at $\sim10$au/outer disc at $\sim100$au case . 

\subsection{avalanches in "extreme'' debris discs?} \label{extreme}

\subsubsection{negligible post-avalanche luminosity drop} \label{drop}

As discussed in Sec.~\ref{intro}, avalanches have been recently invoked as a possible mechanism for explaining the rapid luminosity drops observed in some bright debris discs, under the assumption that, \emph{after} the passage of the avalanche, the remaining main disc would be left depleted from a fraction of its $s\leq1$\,mm grains, which have been "sandblasted" by the avalanche's passage, and would thus become less luminous than what it was before the avalanche.
Our simulations do indeed show a slight luminosity drop (as compared to the pre-avalanche level), in the wake of an avalanche, but this decrease remains limited to less than 1\% in the best case (dense "warm" disc case, see Fig.~\ref{fluxinnerd}), i.e., much less than the luminosity increase due to the avalanche. Moreover this luminosity drop is also limited in time, never lasting much longer than the avalanche itself. We thus conclude that this "sandblasting" eroding effect leaves a very limited imprint on the disc's luminosity, and that the brightness of a post-avalanche disc is very close to the one it had prior to the passing of the avalanche.
The only way to observe a luminosity drop in the context of an avalanche would be to start observing the system \emph{during} an avalanche, more exactly right after the avalanche-induced luminosity peak. In this case the system would indeed appear to become somehow fainter, but this would imply catching the system at a very specific moment during an avalanche's unfolding, and it cannot be considered a generic explanation.

Collisional avalanches could of course explain short term luminosity \emph{increases} in some extreme discs, provided that a big enough object has been shattered. However, even for the very large $m_{\rm{init}}$ considered in our simulations, we do not obtain photometric variations that exceed 10\%. Of course, higher excesses could be reached for even larger initially broken-up objects, but, as underlined in the previous section, the break-up of such larger objects are even less likely to occur.

\subsubsection{vapourized matter}

Another problem is that, for the warm discs that make up most of these extreme systems, avalanches should imply grain-grain collisions at speeds high enough to at least partially vaporize dust grains in the outer main disc. 
For silicates and carbon dust grains \citet{cze2007} estimate that the threshold impact velocity for vapourizing some fraction of a target grain is around $\sim$20km.s.$^{-1}$. However, quantifying the fraction of the target's mass that is vapourized relative to the fraction that is fragmented is a difficult task that requires addressing complex issues about the micro-physics and chemistry of collisions that exceed the scope of the present paper. If we take as a reference the simplified prescription of \citet{cze2007}, and consider, in addition, that a typical avalanche-fueling collision involves an unbound grain hitting a $\sim$5 times bigger target, we find that the target mass that is vapourized becomes comparable to the one converted into solid fragments for velocities approaching $\sim$70km.s$^{-1}$ (see Figs.2 \& 3 and Appendix A of Czechowski \& Mann, 2007). Such very high velocities are in fact reached, for our warm disc case, by $\beta\geq3$ grains, produced at 1\,au by the initial breakup, when they impact bodies in the main disc at 5-12\,au. Even if such $\beta\geq3$ grains do not make up the majority of the initially released unbound population (which peaks around $\beta=5$, see Fig.\ref{betafig}) they do, however, represent a non-negligible fraction of it.

As a consequence, a sizable proportion of the collisions needed to fuel the avalanche will likely produce clouds or plumes of vapour instead of new debris that can further the chain reaction. Eventually, such vapour plumes are expected to recondense into solid matter through a nucleation-growth-quenching process \citep{john2012}, but this condensation process might be much more inefficient in the present case, where the gas density and pressure are likely to be much lower than in the case of large impacts on rocky planets considered by \citet{john2012}. In addition, even if it is succesfull this recondensation process is likely to form a relatively monosize population of spherules in the 100--500$\mu$m range \citep{john2012}, much larger than the submicrometric unbound grains the avalanche's propagation relies upon.
As a consequence, even though quantifying the impact of this vapourization/recondensation process on the avalanche mechanism is a very complex task, it is safe to conclude that our simulations are very likely to overestimate avalanche strength for warm discs.

\subsubsection{TYC 8241 2652 1}

The individual system for which the avalanche scenario has been set forth with most insistence is TYC 8241 2652 1, arguably the most "extreme" extreme disc, which had a very high fractional luminosity ($f\sim$0.11) until the end of 2008 but suddenly (in $\sim$1\,year) became more than one order of magnitude fainter at 10 and $20\mu$m \citep{melis2012}, and has remained at this level since then. 

In the light of the present simulations, we believe that the avalanche scenario can probably be ruled out as a potential explanation for this system's unusual behaviour. First and foremost, we find that the "sandblasting" luminosity drop in the wake of an avalanche is extremely limited, orders of magnitude lower than the factor 30 decrease observed for TYC 8241 2652 1. And even if we consider the unlikely assumption of a "giant" avalanche produced by the breakup of an extremely large object that could have a much stronger sandblasting effect, we would still be unable to explain the photometric evolution of the TYC 8241 2652 1 system \emph{before} the 2009/2010 luminosity drop, i.e., a seemingly constant photometric level for more than 30 years \citep{melis2012} which is impossible to reconcile with the short-lived luminosity peak triggered by an avalanche (see Fig.\ref{fluxinner}).

Another, maybe even more fundamental argument against the avalanche scenario is that TYC 8241 2652 1 is a low mass K star ($L=0.7L_{\odot}$) for which only a very limited range of grain sizes, around $s\sim$0.2$\mu$m, can reach $\beta\geq0.5$ and have unbound orbits (Fig.\ref{betafig}). This leaves much too little "fuel" for creating and propagating an avalanche. This problem is compounded by the fact that this disc's brightness before the luminosity drop, i.e., $f\sim$0.11, combined to its estimated radial distance, $\sim$0.4\,au, makes it optically thick \citep{melis2012}, meaning that stellar radiation is reduced inside the disc, thus further lowering the $\beta$ values of small grains.

\subsection{possible limitations} \label{limit}

Even if our main results and conclusions are probably relatively robust, we are aware that, because of the time-consuming character of the LIDT-DD runs, we could not explore the full parameter space linked to the complex avalanche mechanism. One of these parameters is the location $r_{\rm{init}}$ of the dust-releasing breakup event, which was fixed for both cold (at $10\,$au) and warm (at $1\,$au) disc cases. However, because high-$\beta$ grains quickly reach their asymptotic radial velocity (see Fig.8 of GAT07), the effect of changing $r_{\rm{init}}$ is relatively straightforward to predict:  lower values of $r_{\rm{init}}$ will change the avalanche grain velocity according to $v\propto r_{\rm{init}}^{-0.5}$ while larger values of $r_{\rm{init}}$ will prevent the released grains from reaching high-enough velocities. Likewise, we have not explored the spatial structure of the outer disc, which, apart from being scaled by a factor 10 between the cold and warm cases, remains fixed in terms of its radial slope and relative width $\Delta r/r$. Taking wider discs could for instance increase the value of the radial optical depth $\tau_{r}$ for the same value of $\tau_{\perp}$, which could in principle lead to stronger avalanches by increasing the impact probability of disc-crossing outbound grains without increasing too much the fraction of "native" unbound grains that could weaken the avalanche (see Sec.\ref{densdep}). However, the disc we have considered is already very wide, with a width of $\Delta r/r=0.8$ that is relatively large compared to the more narrow belt-like configurations that seem to be inferred for a large fraction of debris discs \citep{wyat07b,2015Ap&SS.357..103W,2017arXiv170308560K}. As a consequence, our results should probably be considered as a best-case scenario from the point of view of the system's spatial structure (release location and outer disc width). 
Another parameter is the dynamical excitation of the main outer disc. One would maybe expect dynamically "colder" discs to be more favourable to avalanche propagation for the dense $\tau_{\perp}$$\sim$$10^{-2}$ cases, because the lower collisional activity should lead to a lower level of avalanche-hindering "native" unbound grains in the main disc. However, test simulations with lower $<e>$ values did not lead to significant differences in terms of avalanche amplitude. This is mainly because the avalanche-favouring depletion of native unbound grains is compensated by the avalanche-hindering depletion of small \emph{bound} grains (which are the main mass reservoir for fueling the avalanche) that comes with a lower dynamical excitation. This depletion of small bound grains in dynamically cold discs is a classical result that has been identified by \citet{theb08}.

Other potential issues are related to the LIDT-DD code itself, in particular the unavoidably finite resolution of the spatial grid it uses to compute collisions. Ideally, the size of each collision "cell", within which all tracers are considered to mutually collide, should not be much larger than the spatial extension of the propagating avalanche, in order to avoid an artificial dilution effect. In all runs, the azimuthal width of the collision cell is $36\,$ degrees, which is in fact slightly more than the azimuthal extension of the avalanche "arm" as it enters the main disc, which is of the order of $\sim$20--25\,degrees. We thus expect the presence of a some numerical dilution effect, but it should remain relatively moderate and should not drastically change the measured avalanche amplitude and duration

Maybe the most critical issue is that of the amount of small dust released by the initial breakup event. How realistic is it to assume that the shattering of a 40-300\,km-sized planetesimal will release this large amount of sub-micrometre dust and that the post-breakup fragment size distribution can be extended in a self-similar way down to $0.02\mu$m grains (see Table.1)? To our knowledge, there is no reliable experimental or theoretical information regarding the size distribution of the smallest fragments produced after large fragmenting impacts, especially for sizes more then 8 orders of magnitude smaller than the shattered object. One of the few attempts at addressing this problem in the context of debris discs is that of \citet{krij14} and \citet{theb16}, who looked at it from the perspective of surface energy conservation and found that there should indeed be a minimum physical fragment size, which could in some cases be $\geq1\,\mu$m. However, for the case of impacts on large 40-300\,km-sized bodies, this theoretical minimum size should be much lower than the $0.02\mu$m smallest grains considered in the present runs. But the fact that there is no energy-conservation impossibility for the presence of sub-micron fragments does not mean that such fragments will necessarily be produced. This is a very difficult issue that clearly exceeds the scope of the present study, but, here again, our prescription is probably a best case scenario.

\section{Summary and Conclusion}\label{ccl}

A collisional avalanche is a complex mechanism, set-off by the sudden release of a large amount of small grains on unbound orbits (following, for instance, the breakup of a large planetesimal), which then enter at high velocity a debris disc further away from the star, triggering there a collisional chain-reaction producing amounts of dust potentially greatly exceeding the initially released population.

We investigate this mechanism using for the first time a fully self-consistent code, LIDT-DD, coupling the dynamical and collisional evolution of the system. We quantify also for the first time the photometric signature of avalanches, and investigate if they could be responsible for the short-term luminosity variations recently observed in some very bright debris discs. We consider an avalanche-favouring luminous A6V star, and two set-ups: a "cold disc" case with a dust release at 10\,au and an outer disc extending from 50 to 120\,au, and a "warm disc" case scaled down by a factor of 10 (release at 1\,au and 5-12\,au outer disc), taking the outer disc's density (parameterized by its vertical optical depth $\tau_{\perp}$) as a free parameter.

Our main results can be summarized as follows:
\begin{itemize}
\item 
We confirm the results of Grigorieva et al.(2007), and find that avalanches can be set off in outer discs with $\tau_{\perp}\gtrsim$10$^{-3}$. We show, in addition, that this result holds for both "cold" and "warm" discs. 
\item
To reach a local 100\% contrast on resolved scattered light images, the dust mass released by the initial breakup event has to be very large, i.e., $m_{\rm{init}-c}$$\sim$5$\times 10^{22}\,$g (the mass of a $\sim$150\,km body) for the cold disc case, and  $m_{\rm{init}-w}$$\sim$10$^{20}\,$g (the mass of a $\sim$20\,km body) for a warm disc.
\item
The integrated photometric excess of avalanches is more limited, not exceeding 10\% at any wavelength, even for the high $m_{\rm{init}-c}$ and $m_{\rm{init}-w}$ considered here. Furthermore, for the cold disc case, this excess becomes negligible at the long wavelengths were the 50-120\,au disc is the brightest, probably making avalanches very difficult to observe in unresolved Kuiper belt-like cold discs. The situation is more favourable for warm discs, where the avalanche excess peaks at the 10--20$\mu$m wavelengths where the main disc itself is at its brightest. In this mid-IR domain, photometric variations of $\sim$5--10\% should be within the reach of current instruments.
\item
The duration $t_{\rm{aval}}$ of an avalanche is of the order of $0.3\,t_{\rm{orb}-in}$, where $t_{\rm{orb}-in}$ is the orbital period at the inner edge of the main outer disc. This is relatively short but still roughly one order of magnitude larger than the time it would take for breakup-released unbound grains to simply cross the outer disc without colliding.
\item
Contrary to what could be expected from earlier simplified prescriptions, the avalanche strength does not significantly increase when increasing the outer disc's density beyond $\tau_{\perp}\sim$2$\times10^{-3}$. In terms of the \emph{relative} contrast between the avalanche and the main disc, $\tau_{\perp}$  values of a few $10^{-3}$ are probably the optimum case.
This is mainly because, for higher $\tau_{\perp}$ values, the outer disc already has a significant population of unbound grains \emph{before} the avalanche's passage. These "native" unbound grains weaken the avalanche by creating a  dense background  that reduces the relative excess density due to avalanche grains, but also by preemptively eroding the main disc from its population of small bound grains ($s\sim$1--10$\mu$m) that makes up most of the target reservoir for fueling the avalanche propagation. 
\item
We observe a slight luminosity deficit, as compared to the pre-avalanche level, \emph{after} the passage of the avalanche, due to the erosion of $\sim$1--100$\mu$m targets in the disc, but this "sandblasting" effect remains very limited, less than 1\,\%, and short-lived, its duration never greatly exceeding that of the avalanche itself.
\item
This very weak post-avalanche luminosity deficit makes avalanches an unlikely explanation for the much larger photometric drops observed in some very bright "extreme" debris discs, like the $\sim$50\,\% decrease for ID 8 or, even more problematic, the factor 30 luminosity fall of TYC 8241 2652 1. For the latter system, an additional argument against avalanches is that the central star is not luminous enough to place a significant fraction of grains on unbound orbits (a prerequisite for setting off and fueling an avalanche). 
\item
The likelihood for witnessing an avalanche in a given system crucially depends on 2 parameters: 1) the outer disc density, which has to be such as $\tau_{\perp}$  is of the order of a few $10^{-3}$, and 2) the average time $t_{\rm{break}}$ between two avalanche-triggering planetesimal breakups, the probability of witnessing an avalanche being $\sim t_{\rm{aval}}/t_{\rm{break}} $. 
\item
We estimate that the "ideal" case would thus be a system with a double-disc structure, with an inner belt (located at $\sim$1\,au or  $\sim$10\,au depending on the "cold" or "warm" disc case) of fractional luminosity $f$$\gtrsim$10$^{-4}$, which produces a high enough rate of avalanche-triggering breakups, and an outer disc with $\tau_{\perp}$$\sim2\times10^{-3}$, located at $\sim$5--10 times the location of the inner belt, into which the avalanche propagates. Cold disc systems would then be the most favourable to the detection of features on resolved images in scattered-light, because of their larger scale, while warm discs would be more suited to the detection of photometric variations in the mid-IR, because the avalanche-induced excess peaks at wavelengths where the disc itself is at its brightest.
Such double-belt configurations seem to have been identified for several observed discs \citep{geil17}, and will be investigated in future studies. 

\end{itemize}

\begin{acknowledgements}
The authors thank Rik Van Lieshout for fruitful discussions and an anonymous reviewer for an insightful referee.
QK acknowledges funding from STFC via the Institute of Astronomy, Cambridge, consolidated grant.
PT dedicates this paper to the memory of Andr\'e Brahic.

\end{acknowledgements}

{}

\clearpage


\begin{thebibliography}{}

%
\bibitem[Artymowicz(1997)]{arty97} Artymowicz, P.\ 1997, Annual Review of Earth and Planetary Sciences, 25, 175
%
\bibitem[Augereau et al.(1999)]{auge99}Augereau, J. C., Lagrange, A. M., Mouillet, D., Papaloizou, J. C. B., \& Grorod, 1999, A\&A, 358, 557
%
\bibitem[Benz \& Asphaug(1999)]{benz99} Benz, W., \& Asphaug, E.\ 1999, \icarus, 142, 5 
%
\bibitem[Beust et al.(2001)]{2001A&A...366..945B} Beust, H., Karmann, C., \& Lagrange, A.-M.\ 2001, \aap, 366, 945 
%
\bibitem[Bonsor et al.(2013)]{bonsor2013} Bonsor, A., Augereau, J.C., Thebault, P., 2013, A\&A, 548, 104
%
\bibitem[Bottke et al.(2005)]{bott05}  Bottke, W., Durda, D., Nesvorny, D., Jedicke, R., Morbidelli, A.,Vokourhlicky, D., Levison, H., 2005, Icarus, 179, 63
%
\bibitem[Charnoz \& Taillifet(2012)]{char12} Charnoz, S., Taillifet, E., 2012, ApJ, 753, 119
%
\bibitem[Chen et al.(2014)]{2014ApJS..211...25C} Chen, C.~H., Mittal, T., Kuchner, M., et al.\ 2014, \apjs, 211, 25 
%
\bibitem[Czechowski \& Mann(2007)]{cze2007} Czechowski, A., Mann, I., 2007, ApJ, 660, 1541
%
\bibitem[Dermott et al.(2002)]{derm02} Dermott, S., Kehoe, T., Durda, D., et al., 2002, in Proc., Asteroids, Comets, Meteors, ed B. Warmbein, 319
%
\bibitem[Dohnanyi(1969)]{dohn69} Dohnanyi J.S., 1969, JGR 74, 2531
%
\bibitem[Draine(2003)]{drai03} Draine, B.~T.\ 2003, \araa, 41, 241
%
\bibitem[Farinella et al.(1998)]{fari98} Farinella, P., Vokroulicky, D., Hartmann, W., 1998, Icarus, 132, 378 
%
\bibitem[G{\'a}sp{\'a}r et al.(2013)]{gasp13} G{\'a}sp{\'a}r, A., Rieke, G.~H., \& Balog, Z.\ 2013, ApJ, 768, 25 
%
\bibitem[Geiler \& Krivov(2017)]{geil17} Geiler, F., \& Krivov, A.~V.\ 2017, \mnras, 468, 959 
%
\bibitem[Genda et al.(2015)]{genda2015} Genda, H., Kobatashi, H., Kokubo, E., ApJ, 810, 136
%
\bibitem[Grigorieva et al.(2007)]{grig07} Grigorieva, A., Artymowicz, P., Thebault, P., 2007, A\&A, 461, 537
%
%\bibitem[Jackson \& Wyatt(2012)]{jack12} Jackson, A.~P., \& Wyatt, M.~C.\ 2012, \mnras, 425, 657 
%
\bibitem[Jackson et al.(2014)]{jack14} Jackson, A.~P., Wyatt, M.~C., Bonsor, A., \& Veras, D.\, 2014, MNRAS, 440, 375 
%
\bibitem[Johnson \& Melosh(2012)]{john2012} Johnson, B.~C., Melosh, H. J. \ 2012, Icarus, 217, 416 
%
\bibitem[Kennedy \& Wyatt(2013)]{kennedy2013} Kennedy, G.M., Wyatt, M.C., 2013, MNRAS, 433, 2334
%
\bibitem[Krijt \& Kama(2014)]{krij14} Krijt, S, Kama, M., 2014, A\&A, 566, 2
%
\bibitem[Kral et al.(2013)]{kral13} Kral, Q.; Thebault, P.; Charnoz, S., 2013, A\&A, 558, A121
%
\bibitem[Kral et al.(2015)]{kral15} Kral, Q.; Thebault, P.; Augereau, J.-C., Boccaletti, A.,Charnoz, S., 2015, A\&A, 573, A39
%
\bibitem[Kral et al.(2017a)]{2017arXiv170308560K} Kral, Q., Clarke, C., \& Wyatt, M.\ 2017, arXiv:1703.08560 
%
\bibitem[Kral et al.(2017b)]{2017arXiv170302540K} Kral, Q., Krivov, A.~V., Defrere, D., et al.\ 2017, arXiv:1703.02540 
%
\bibitem[Krivov(2010)]{kriv10} Krivov, A.~V.\ 2010, Research in Astronomy and Astrophysics, 10, 383 
%
\bibitem[Leinhardt \& Stewart(2012)]{lein12} Leinhardt, Z.~M., \& Stewart, S.~T.\ 2012, ApJ, 745, 79 
%
\bibitem[Levison et al.(2012)]{levison12} Levison, H.~F., Duncan, M.~J., \& Thommes, E.\ 2012, AJ, 144, 119 
%
\bibitem[Lisse et al.(2009)]{liss09} Lisse, C.~M., Chen, C.~H., Wyatt, M.~C., et al.\ 2009, \apj, 701, 2019 
%
\bibitem[L{\"o}hne et al.(2008)]{lohn08} L{\"o}hne, T., Krivov, A.~V., \& Rodmann, J.\ 2008, \apj, 673, 1123 
%
\bibitem[Lyra \& Kuchner(2013)]{lyra2013} Lyra, W., Kuchner, M., 2013, Nature, 449, 184
%
\bibitem[Matthews et al.(2014)]{matthews2014}  Matthews, B. C.; Krivov, A. V.; Wyatt, M. C.; Bryden, G.; Eiroa, C., 2014, in Protostars and Planets VI, Henrik Beuther, Ralf S. Klessen, Cornelis P. Dullemond, and Thomas Henning (eds.), University of Arizona Press, Tucson, 521-544
%
\bibitem[Melis et al.(2012)]{melis2012} Melis, C., Zuckerman, B., Rhee, J., Song, I., Murphy, J., Bessel, M., 2012, Nature, 487, 75
%
\bibitem[Melis(2016)]{melis2016} Melis, C., 2016, 
%
\bibitem[Meng et al.(2014)]{meng2014} Meng, H., Su, K., Rieke, G., et al., 2014, Science, 345, 1032
%
\bibitem[Meng et al.(2015)]{meng2015}  Meng, H., Su, K., Rieke, G., et al., 2015, ApJ, 805, 77
%
\bibitem[Mouillet et al.(1997)]{moui97} Mouillet, D.; Larwood, J. D.; Papaloizou, J. C. B.; Lagrange, A. M., 1997, MNRAS, 292, 896
%
\bibitem[Nesvold et al.(2013)]{nesvold13} Nesvold, E.~R., Kuchner, M.~J., Rein, H., \& Pan, M.\ 2013, ApJ, 798, 83 
%
\bibitem[Nesvold \& Kuchner(2015)]{nesvold15} Nesvold, E.~R., Kuchner, M.~J., \ 2015, ApJ, 777, 144 
%
\bibitem[Nesvold et al.(2016)]{nesvold16} 	Nesvold, Erika R.; Naoz, Smadar; Vican, Laura; Farr, Will M., \ 2016, ApJ, 826, 19 
%
\bibitem[Pearce \& Wyatt(2015)]{2015MNRAS.453.3329P} Pearce, T.~D., \& Wyatt, M.~C.\ 2015, \mnras, 453, 3329 
%
\bibitem[Reche et al.(2009)]{reche09} Reche, R.; Beust, H.; Augereau, J.-C., 2009, A\&A, 463, 661
%
\bibitem[Rieke et al.(2008)]{rieke08} Rieke, G. H.; Blaylock, M.; Decin, L.; et al., 2008, AJ, 135, 2245
%
\bibitem[Schneider et al.(2014)]{2014AJ....148...59S} Schneider, G., Grady, C.~A., Hines, D.~C., et al.\ 2014, \aj, 148, 59 
%
\bibitem[Sezestre et al. (2017)]{seze17} Sezestre, E., Augereau, J.-C., Boccaletti, A., Thebault, P., 2017, A\&A, aceepted for publication, arXiv:1707.09761
%
\bibitem[Song et al.(2005)]{song05}  Song, I., Zuckerman, B., Weinberger, A., Becklin, E., 2005, Nature, 436, 363
%
\bibitem[Takasawa et al.(2011)]{taka11} Takasawa, S., Nakamura, A. M., Kadono, T., et al., 2011, ApJ, 733, L39
%
\bibitem[Takeuchi \& Artymowicz(2001)]{take01} Takeuchi, T., Artymowicz, P., 2001, ApJ, 557, 990
%
\bibitem[Thebault(2009)]{theb09} Thebault, P., 2009, A\&A, 505, 1269
%
\bibitem[Thebault(2012)]{theb12a} Thebault, P., 2012, A\&A, 537, 65
%
\bibitem[Thebault(2016)]{theb16} Thebault, P., 2016, A\&A, 587, 88
%
\bibitem[Thebault et al.(2012)]{theb12b} Thebault, P.; Kral, Q.; Ertel, S. A\&A, 2012, 547, 92
%
\bibitem[Thebault et al.(2010)]{theb10} Thebault, P.; Marzari, F.; Augereau, J.-C., A\&A, 2010, 524, 13
%
\bibitem[Thebault \& Wu(2008)]{theb08} Thebault, P.; Wu, Y., A\&A, 2008, 481, 713
%
\bibitem[Wyatt(2006)]{wyat06} Wyatt, M. C.; 2006, ApJ, 639, 1153
%
\bibitem[Wyatt et al.(2007a)]{wyat07a} Wyatt, M. C.; Smith, R.; Greaves, J. S.; Beichman, C. A.; Bryden, G.; Lisse, C. M., 2007, ApJ, 658, 569
%
\bibitem[Wyatt et al.(2007b)]{wyat07b} Wyatt, M. C.; Smith, R.; Su, K. Y. L.; Rieke, G. H.; Greaves, J. S.; Beichman, C. A.; Bryden, G , 2007, ApJ, 663, 365
%
\bibitem[Wyatt(2008)]{wyat08} Wyatt, M.~C.\ 2008, ARA\&A, 46, 339 
%
\bibitem[Wyatt et al.(2015)]{2015Ap&SS.357..103W} Wyatt, M.~C., Pani{\'c}, O., Kennedy, G.~M., \& Matr{\`a}, L.\ 2015, \apss, 357, 103 
%
\bibitem[Wyatt \& Jackson(2016)]{wyatt2016} Wyatt, M. C., Jackson,  A.P., 2016, SSSr, 205, 231
%
\bibitem[Zuckerman \& Song(2012)]{zuck12} 	Zuckerman, B.; Song, Inseok, 2012, ApJ, 758, 77
%
%
\end{thebibliography}
\end{document}